\RequirePackage[2020-02-02]{latexrelease}
\UseRawInputEncoding

\documentclass[reprint,amsmath,amssymb,aps]{revtex4-2}

\usepackage{graphicx}
\usepackage{subfig}

\begin{document}

\title{Translocation of a single Arg$_9$ peptide across a DOPC/DOPG(4:1) model membrane using the weighted ensemble method}

\author{Seungho Choe}
\email{schoe@dgist.ac.kr}
\affiliation{Department of Energy Science \& Engineering, Daegu Gyeongbuk Institute of Science \& Technology (DGIST), Daegu 42988, South Korea}
\affiliation{Energy Science \& Engineering Research Center, Daegu Gyeongbuk Institute of Science \& Technology (DGIST), Daegu 42988, South Korea}


\begin{abstract}
It is difficult to observe a spontaneous translocation of cell-penetrating peptides(CPPs) within a short time scale (e.g., a few hundred ns) in all-atom molecular dynamics(MD) simulations because the time required for the translocation of usual CPPs is on the order of minutes or so.
In this work, we report a spontaneous translocation of a single Arg$_9$(R9) across a DOPC/DOPG(4:1) model membrane within an order of a few tens ns scale by using the weighted ensemble(WE) method. We identify how water molecules and the orientation of Arg$_9$ play a role in translocation. We also show how lipid molecules are transported along with Arg$_9$. 
 In addition, we present free energy profiles of the translocation across the membrane using umbrella sampling and show that a single Arg$_9$ translocation is energetically unfavorable.
We expect that the WE method
can help study interactions of CPPs with various model membranes within MD simulation approaches.

\end{abstract}

\maketitle

\section{Introduction \label{sec1}}

Cell-penetrating peptides(CPPs) have been extensively studied for several decades because of their capability to transport various cargoes into cells \cite{Guidotti2017,Walrant2017}. 
Multiple factors affect the transport mechanisms of CPPs, e.g., the concentration of CPPs and the properties of the membrane \cite{Pisa2014,Ruseska2020}.
One of the most common difficulties in studying CPPs is that the translocation mechanisms of different CPPs are not the same, and most CPPs can have more than a single
pathway depending on the experimental conditions \cite{Madani2011}.

Molecular dynamics (MD) simulations have been a valuable tool for revealing the mechanical and functional properties of CPPs and their interactions with lipid bilayers \cite{Herce2007,Herce2009,Yesylevskyy2009}; however, the transport mechanism of CPPs and interactions with lipids are still unclear within MD simulation approaches. One of the issues in MD simulations is that it is difficult to observe the spontaneous translocation of CPPs within a few hundred ns time scale because it usually takes minutes or so for the translocation of CPP in experiments \cite{Zorko2005}. Therefore, people have been using biased simulations, such as the umbrella sampling \cite{Pourmousa2013,Sun2014,Tesei2017,Yao2019,Choong2021} and steered MD simulations \cite{Akhunzada2017}, to study CPPs and their interactions with lipid bilayers during translocation. The umbrella sampling is very popular for obtaining free energy barriers between CPPs and lipid bilayers. However, there could be artifacts in the umbrella sampling and thus its free energy analysis because the initial conformations are usually generated by biased simulations (e.g., SMD simulations).  

Among various CPPs, arginine (R)-rich peptides have been extensively studied in experiments and simulations because of their effectiveness in translocation \cite{Fretz07,Ruseska2020}. Strong interaction with negatively charged phospholipid heads is the primary mechanism of inserting R-rich peptides into the lipid bilayer. 
In the previous study, we implemented a weighted ensemble (WE) method \cite{Huber1996,Zuckerman2017} in all-atom MD simulations of Arg$_9$(R9) with a DOPC/DOPG(4:1) model membrane \cite{Choe2021}.
The WE method is a very flexible path sampling technique and is easy to implement in any MD package. 
The WE method uses an ensemble of simulation trajectories. Each trajectory is independent and has a statistical weight. The progress (or reaction) coordinate is divided into multiple bins. Trajectories are periodically replicated in bins if there are too few trajectories, while they are pruned in bins if there are too many trajectories \cite{Zuckerman2017}. 
Zuckerman and Chong \cite{Zuckerman2017} gave a detailed review. We used the WESTPA software \cite{westpa,Bogetti2019,Russo2022} that has been widely applied to various systems.
One of the most significant advantages of using the WE method is that one can simulate a system without any biased potential. 
In our previous simulations \cite{Choe2021}, we found the WE method was very effective for studying interactions between Arg$_9$ and the model membrane. However, we couldn't observe a spontaneous translocation of Arg$_9$ across the membrane because Arg$_9$ was stuck in the hydrophobic core of the model membrane and couldn't move for a long time \cite{Choe2021}.

In this study, the previous WE simulation \cite{Choe2021} was continued with a few different boundaries and bin sizes, and we finally observed a spontaneous translocation of Arg$_9$ across the membrane.  
In the following sections, we show how a single Arg$_9$ peptide can translocate across the model membrane using the WE method and how the orientation of Arg$_9$ affects the translocation efficiency. 
We believe that the WE method will help study the translocation of various CPPs and their transport mechanisms.

\section{Methods \label{sec2}}

\subsection*{Conventional MD simulations}

All simulations were performed using the NAMD package \cite{namd} and CHARMM36 force field \cite{charmm}. In addition, we used a pre-equilibrated structure from the previous study \cite{Choe2020}, equilibrated up to 1 $\mu s$, that was also used in the previous WE simulations \cite{Choe2021}. The detailed method regarding the WE simulations is provided in the following subsection.
Here is a recap of the equilibrium simulation: 
Our system contains 4 Arg$_9$, DOPC/DOPG(4:1) membrane, TIP3P water molecules, and ions generated by CHARMM-GUI \cite{charmm-gui}. 
It was well equilibrated before starting the WE simulations. The DOPC/DOPG(4:1) mixture consists of 76 DOPC and 19 DOPG lipids in each layer.
K$^+$ and Cl$^-$ ions were added to each system to make a concentration of 150 mM.
All Arg$_9$ peptides were
initially located in the upper solution and bound to the upper layer during the equilibration.
The NPT simulations were
performed at T = 310K. Temperature and pressure were kept constant using Langevin
dynamics. An external electric field(0.05 V/nm) was applied in the negative z-direction(from CPPs
to the membrane) as suggested in the previous work \cite{Herce2009,Walrant2012} and also in our earlier simulations \cite{Choe2020,Choe2021} to account for the transmembrane potential \cite{Roux2008}. The particle-mesh Ewald(PME) algorithm was used to compute the
electric forces, and the SHAKE algorithm allowed a 2 fs time step during the simulation.
During the equilibration, CPPs were confined in the upper water box, so  there was no interaction between CPPs and the lower leaflet of model membranes. Whenever a CPP was leaving the upper water box, a small force was applied to pull that CPP inside the box. All CPPs were well contacted with the lipid molecules after the 1 $\mu$s long equilibration, and then the WE simulations were performed using this equilibrated system.

\subsection*{Weighted Ensemble (WE) simulations}

We use the WESTPA (The Weighted Ensemble Simulation Toolkit with Parallelization and
Analysis) software package (v2020.05, see https://github.com/westpa/westpa/wiki/Installing-WESTPA)  \cite{westpa,Bogetti2019,Russo2022} to enable the simulation of rare events, for example, the translocation of CPPs across a model membrane. WESTPA is open-source, and its utility has been proven for many problems.  
All WE trajectories are unbiased and used to calculate conditional probabilities or transition rates\cite{Zuckerman2017}. 

To use the WE method in MD simulations, we need to define a progress coordinate, the total number of bins, the number of walkers (child simulations or replicas) in each bin, and a time interval for splitting and combining trajectories \cite{westpa}.
We define the progress coordinate as a distance in the z-direction between the center of mass of phosphorus atoms in the upper leaflet and that of Arg$_9$. After equilibration of the system, an initial distance between phosphorus atoms and Arg$_9$ was measured, and a boundary was set using this initial position and the position of the center of the membrane, for example, [$-$18 \AA ~(the center of membrane), 3 \AA ~(the initial distance)]. Each boundary can have different bin sizes (0.25 \AA ~or 0.1 \AA). For example, we set a smaller bin size when the peptide has difficulty overcoming the free energy barriers. The number of walkers (replicas) in each bin was 5.  Due to the shortage of computing resources, we had to run several WE simulations to combine all the trajectories and observe  the translocation of Arg$_9$ (Table \ref{tab1}). In the current setup (for example, in WE2), replicas can move above the upper boundary ($-$14 ~\AA); however, replicas can not move below the lower boundary ($-$25 ~\AA). Therefore, when one of the replicas reached the lower boundary, the WE simulation was stopped automatically, and we continued the simulation at this position with a new boundary. The time interval for splitting-combining the trajectories (called an iteration in Table \ref{tab1}) during each WE simulation was 5 ps. The total iterations during all the WE simulations were 5080. The progress coordinate was calculated using MDAnalysis \cite{Michaud2011}. 

\begin{table}[ht]
\begin{center}
\caption{A list of WE simulations}\label{tab1}%
\begin{tabular}{@{}cccc@{}}
\hline
simulation no.  & bin boundaries  &  bin size (the smallest) & \ \# iterations \\
\hline
WE1    & [$-$18.0 \AA, 6.0 \AA]   &  0.25 \AA  & 664  \\
WE2    &  [$-$25.0 \AA, $-$14.0 \AA]  &  0.25 \AA & 3218  \\
WE3    & [$-$28.0 \AA, $-$21.0 \AA]  & 0.10 \AA & 488  \\
WE4   & [$-$35.0 \AA, $-$25.0 \AA] & 0.10 \AA  & 514 \\
WE5    & [$-$45.0 \AA, $-$33.0 \AA] & 0.10 \AA   & 196 \\
\hline
\end{tabular}
\end{center}
\end{table}

\subsection*{Umbrella Sampling}

We used umbrella sampling to obtain the PMF along the translocation path. We chose 49 frames among the total of 5080 frames(iterations). Note that the size of each window is not uniform. We used the colvar module for the umbrella sampling\cite{namd}. The distance in the z direction between the center of mass of $C_\alpha$s in Arg$_9$ and that of phosphorus atoms was restrained using the harmonic restraints with a force constant ($k$ = 10.0~ kcal/mol/ \AA$^2$). We sampled simulation data at each window for 140 ns. We discarded the initial 20 ns and analyzed the rest of the data (120 ns) using the WHAM(weighted histogram analysis method) code \cite{Grossfield2022}.

\section{Results \label{sec3}}

\subsection*{Translocation of a single Arg$_9$ peptide was observed within a very short time scale}

We observe that a single Arg$_9$ translocates the model membrane during the WE simulations. In the previous WE simulation, the translocation of Arg$_9$ was not observed because Arg$_9$ was trapped in the hydrophobic core of the membrane for a long time \cite{Choe2021}. In the current simulations, the size of each bin was decreased from 0.25 \AA ~in the previous WE simulation \cite{Choe2021} to 0.10 \AA, which could help Arg$_9$ overcome free energy barriers along the translocation path. The progress coordinate was defined as the distance between the center of mass of one of Arg$_9$s and that of the phosphorus(P) atoms in the upper leaflet (see the Methods section). The other three Arg$_9$ peptides were closely contacted with the upper leaflet during the simulations, and they didn't show any translocation or any meaningful penetration.

\begin{figure}[ht!]%
\centering
\includegraphics[width=8cm,height=6cm]{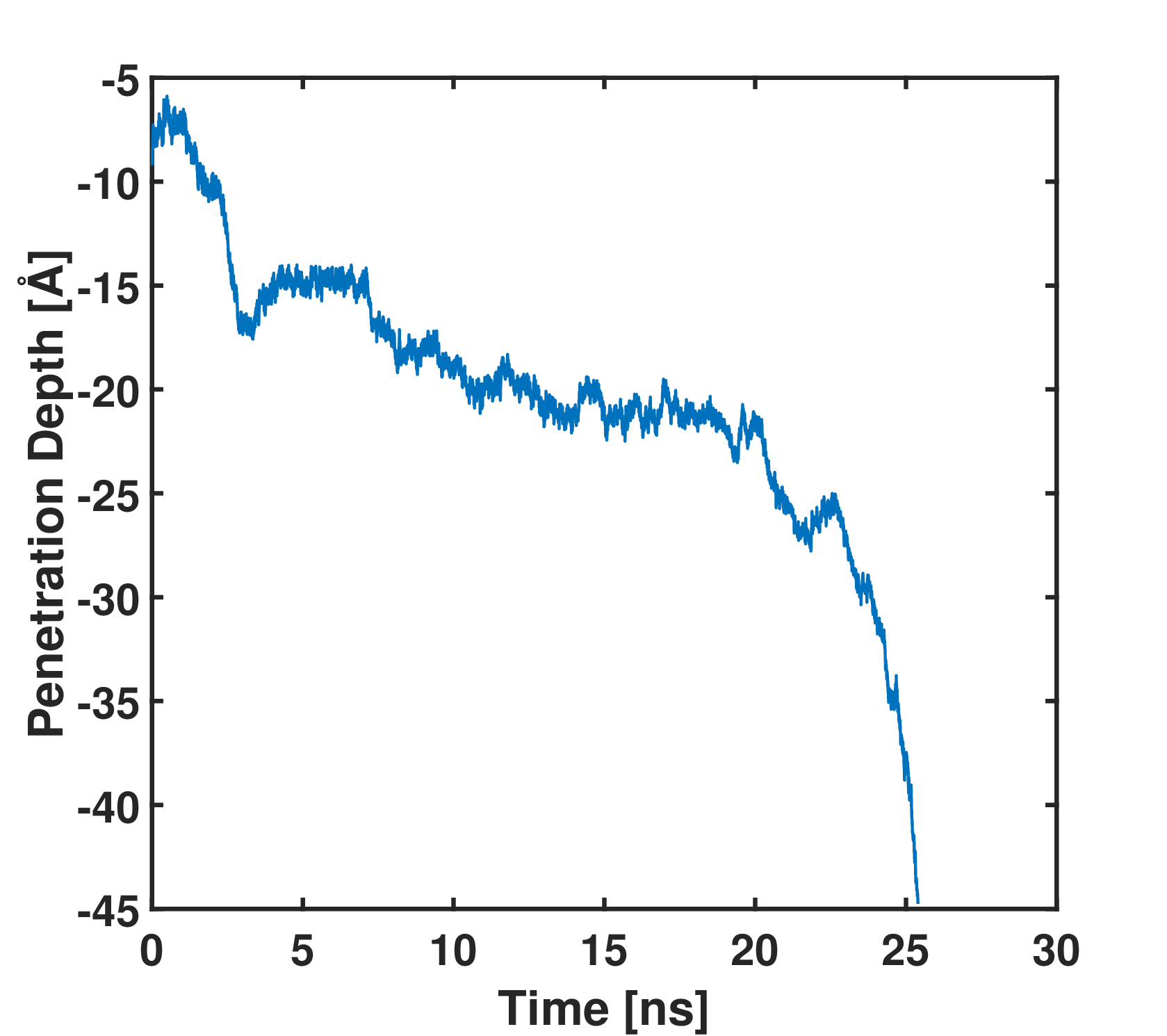}
\caption{The penetration depth vs. the simulation time.}\label{fig1}
\end{figure}

Fig. \ref{fig1}
shows the penetration depth of Arg$_9$ vs. the 
simulation time. 
The total number of splitting-combining trajectories was 5080 in our WE simulations, and this number was converted to the simulation time using 5 ps as the time interval for splitting and combining trajectories (see the Methods section).
We define the penetration depth as the distance between the center of mass of Arg$_9$ and that of the upper leaflet's phosphorus(P) atoms.
Therefore, the negative sign denotes that Arg$_9$ is located below the upper leaflet. 
The maximum penetration depth of Arg$_9$ in the previous WE simulation was about $- 17.6$ \AA ~\cite{Choe2021}, and now it can reach up to $- 45$ \AA ~in the current simulations. 

As shown in the figure, the penetration depth changes rapidly within a very short time scale. Furthermore, the figure shows that Arg$_9$ can translocate the model membrane quickly ($\sim$ 25 ns) since the WE simulation started. 
Therefore, the WE method provides a very effective tool to overcome free energy barriers between Arg$_9$ and the model membrane. Thus the WE method could be used to study interactions between membrane-active peptides(MAPs), including cell-penetrating peptides(CPPs) and antimicrobial peptides(AMPs), and various model membranes within MD simulation approaches.

\begin{figure}[ht!]%
\centering
\subfloat[7.5 ns]{\includegraphics[width=4cm,height=4cm]{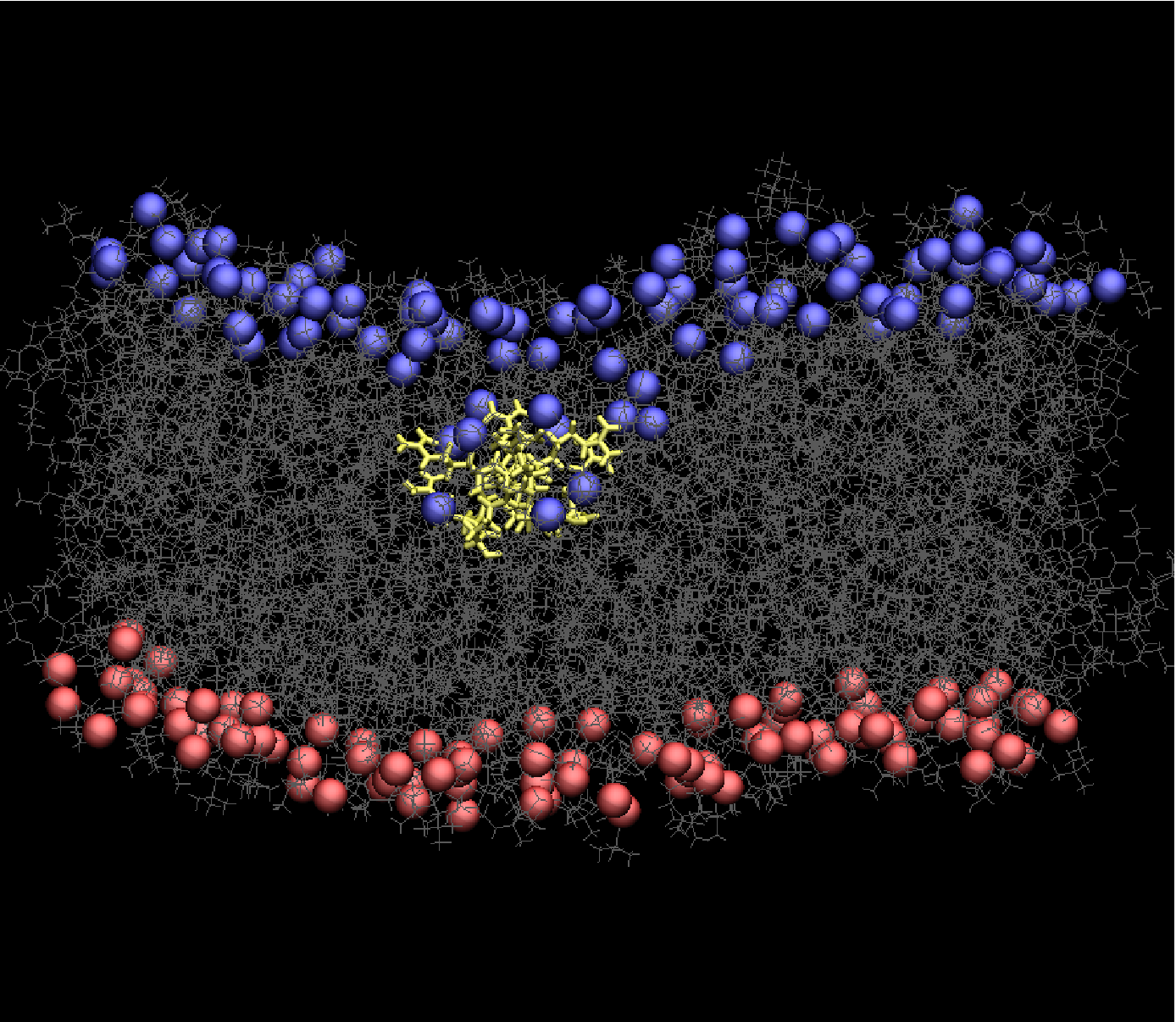}}\vspace{5mm}
\subfloat[15.0 ns]{\includegraphics[width=4cm,height=4cm]{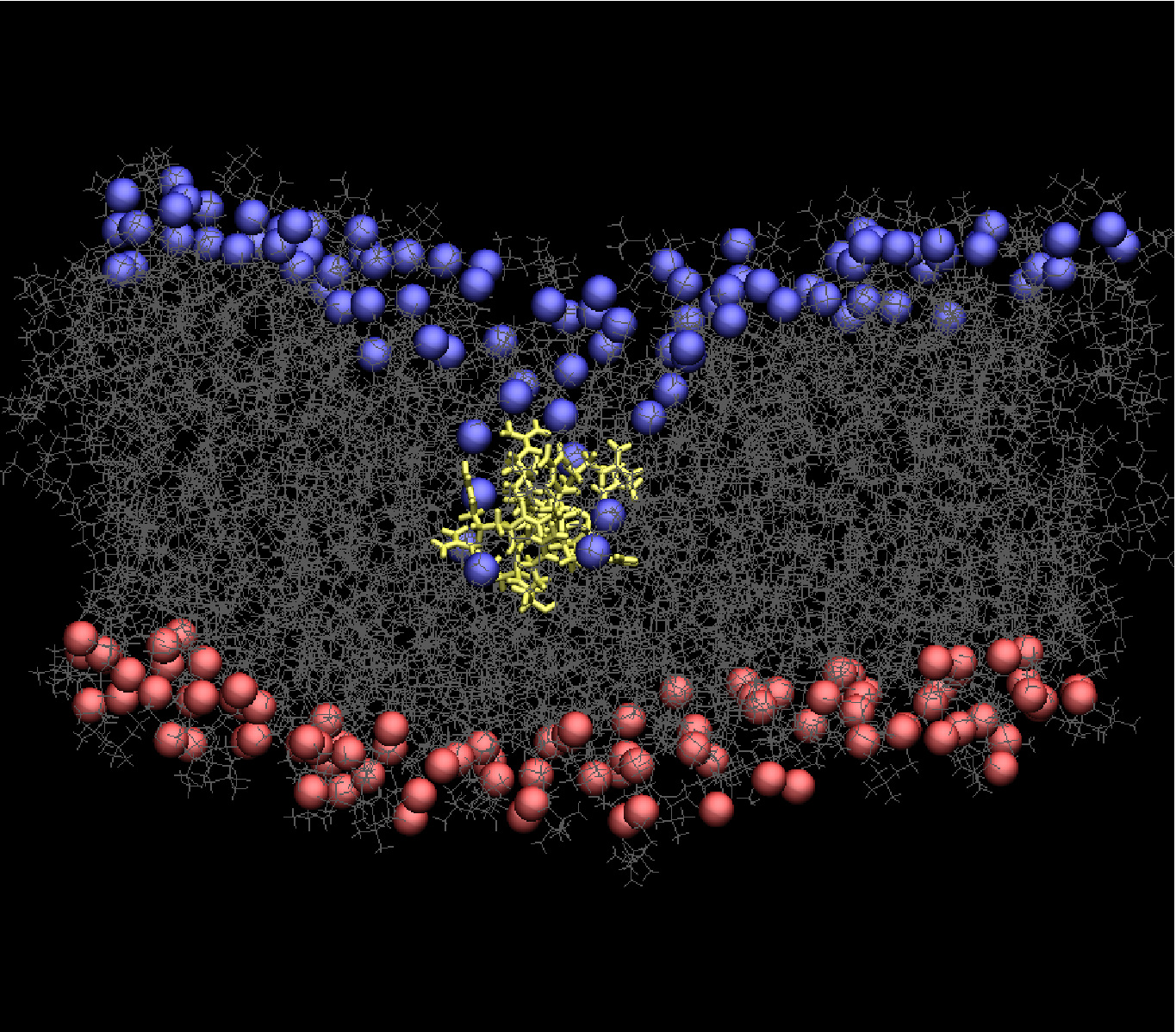}}\vspace{5mm}
\subfloat[22.5 ns]{\includegraphics[width=4cm,height=4cm]{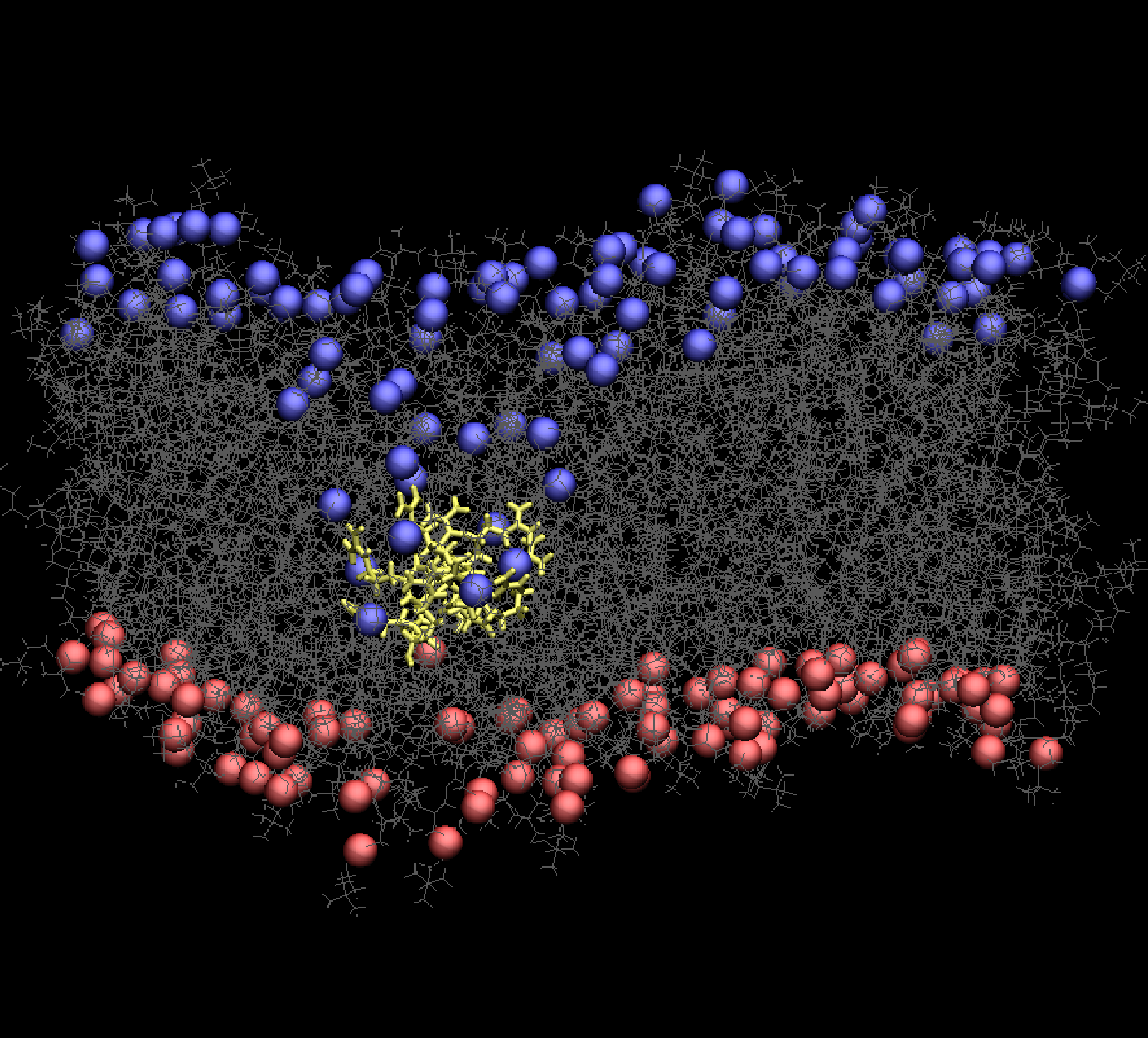}}\vspace{5mm}
\subfloat[the last (25.4 ns)]{\includegraphics[width=4cm,height=4cm]{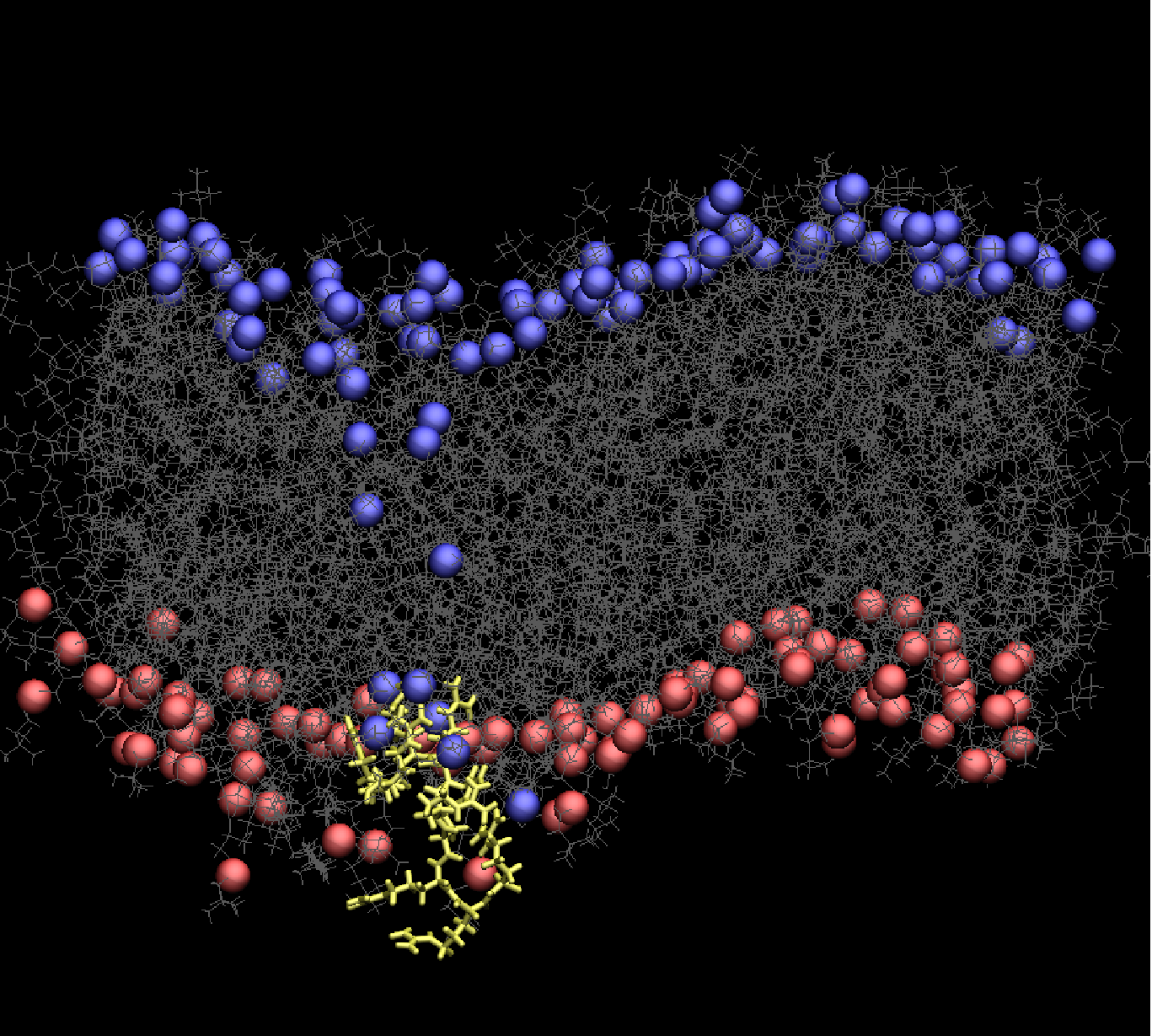}}
\caption{Snapshots at (a) 7.5 ns, (b) 15.0 ns, (c) 22.5 ns, and (d) the last (25.4 ns). (yellow: Arg$_9$, grey: lipids, blue \& red: phosphorus atoms of the upper leaflet \& lower leaflet, respectively)}\label{fig2}
\end{figure}

Fig. \ref{fig2} presents snapshots of Arg$_9$ during the WE simulations. They are snapshots at 7.5 ns, 15.0 ns, 22.5 ns, and the last (25.4 ns). In the figure, yellow shows Arg$_9$, and gray is the lipid molecules in the model membrane (DOPC/DOPG(4:1)). The blue and red dots are phosphorus atoms of the upper and the lower leaflets, respectively. Water molecules, ions, and the other three Arg$_9$s are omitted for clarity. As shown in Fig. \ref{fig2}, the
membrane is deformed when Arg$_9$ penetrates the middle of the membrane. There are two interesting findings during the deformation of the model membrane: The first finding is that a part of the lipids in the upper leaflet moved along with Arg$_9$. The last snapshot showed that these lipids went through the lower leaflet. Later, we will discuss how these lipids are  transported and reorganized in the model membrane. The second finding is the disruption of the lower leaflet both at 22.5 ns and at the last (25.4 ns) shown in the figure. When Arg$_9$ comes to the bottom of the membrane, the lower leaflet deforms drastically, widening of a water pore in the lower leaflet.  

During the translocation, one of the interesting quantities to analyze would be hydrogen bond formation between Arg$_9$ and the lipids molecules.
It has been known that the hydrogen bonding between CPPs and lipids (or water) is critical when CPPs contact the membrane at the beginning of the penetration and when they translocate across the membrane \cite{Rothbard2004,Tang2008,Takechi-Haraya2018,Pei2022}.
Fig. S1 presents the number of hydrogen bonds during the translocation (from WE1 to WE5 simulations). The blue line denotes the number of hydrogen bonds between Arg$_9$ and the lipid molecules (including the phosphate group), while the green line is between Arg$_9$ and only the phosphate group in the lipids. The red line depicts the ones between Arg$_9$ and water. Each point in the figure is an averaged value over 50 ps. The average number of hydrogen bonds between Arg$_9$ and lipids and water is about 20, showing slight fluctuation during the translocation.
This indicates that the strength of hydrogen bonding doesn't affect the efficiency of the translocation. Although the hydrogen bonding between Arg$_9$ and the membrane is vital at the initial penetration stage, our simulation suggests that another factor could be responsible for translocating Arg$_9$ across the membrane. We think water can be another factor that makes the translocation. We will discuss this issue in the following section. 

Another interesting quantity to analyze is the conformational change of Arg$_9$ during the translocation. It has been known that Arg$_9$ stayed at a random coil both in the solution and in the membrane environment \cite{Eiriksdottir2010}. Fig. S2 shows the conformational change of four Arg$_9$s during the WE simulation, identified using the Timeline in VMD \cite{vmd}. The red box corresponds to the only Arg$_9$ that translocated the membrane. During the translocation (Arg$_9$ in the red box), most arginine residues showed a turn(aqua) in the Timeline graph, and this structure remained until the end of the simulation. The other three Arg$_9$s showed a random coil(white) or a combination of both a coil and a turn, and this is because most parts of Arg$_9$s were exposed to water molecules in the upper solution. Our simulation suggests a distinct conformational change (from a random coil to a turn) when Arg$_9$ translocates across the membrane. More translocated trajectories are needed to reveal the conformational changes of Arg$_9$ and their roles for the translocation across the membrane.

\subsection*{Water molecules play a role in translocating Arg$_9$ across the model membrane}

The pore formation process has been well known based on MD simulations of one of the antimicrobial peptides, melittin \cite{Sengupta2008}. According to their results, a P/L(peptide/lipid) ratio above a threshold (P/L=1/64) and aggregation of peptides were necessary for making pores. Although we didn't see the aggregation of Arg$_9$s, our simulation suggests that even a single Arg$_9$ can induce a water pore.

\begin{figure}[ht!]%
\centering
\subfloat[7.5 ns]{\includegraphics[width=4cm,height=4cm]{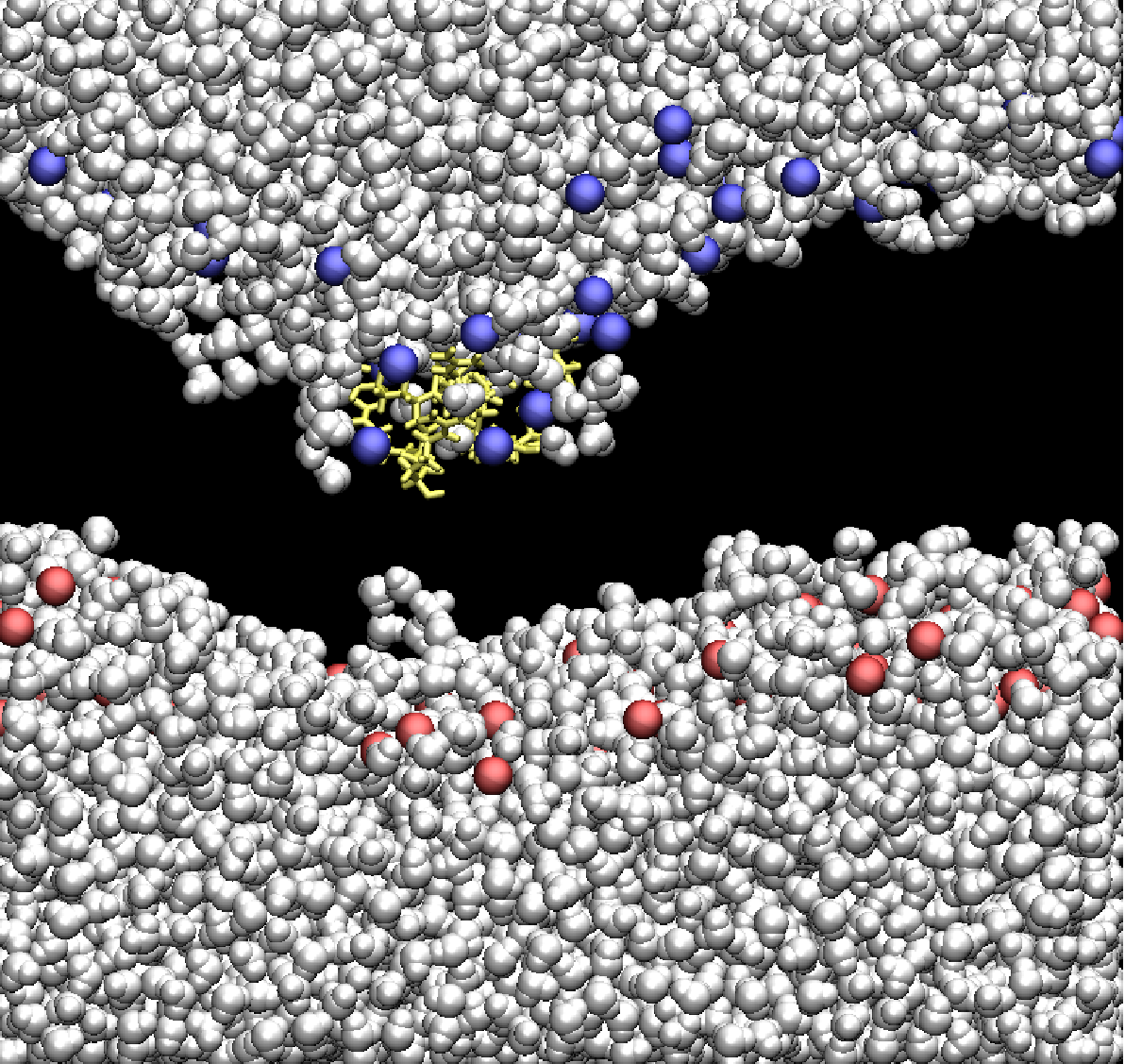}}\vspace{5mm}
\subfloat[15.0 ns]{\includegraphics[width=4cm,height=4cm]{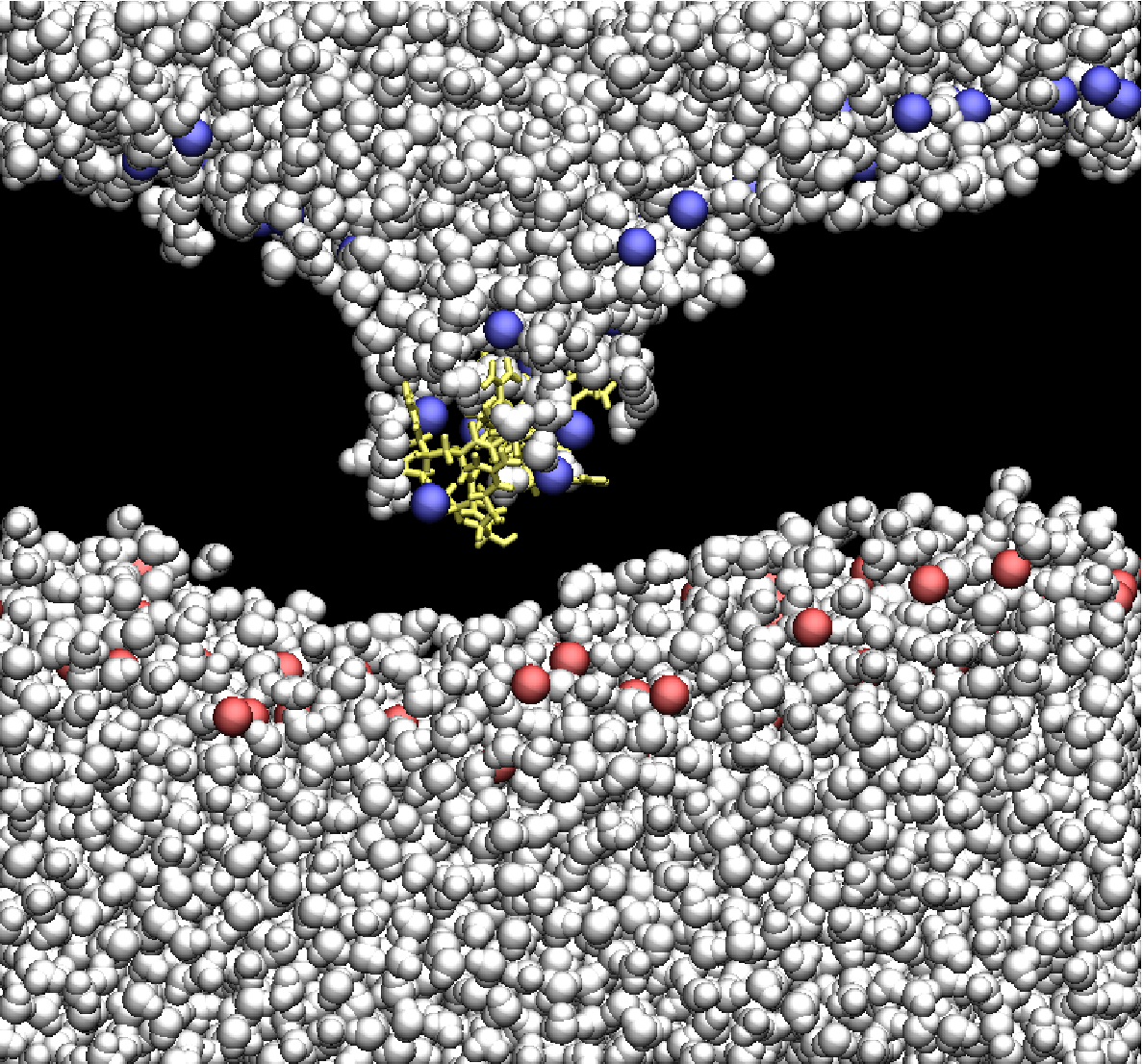}}\vspace{5mm}
\subfloat[22.5 ns]
{\includegraphics[width=4cm,height=4cm]{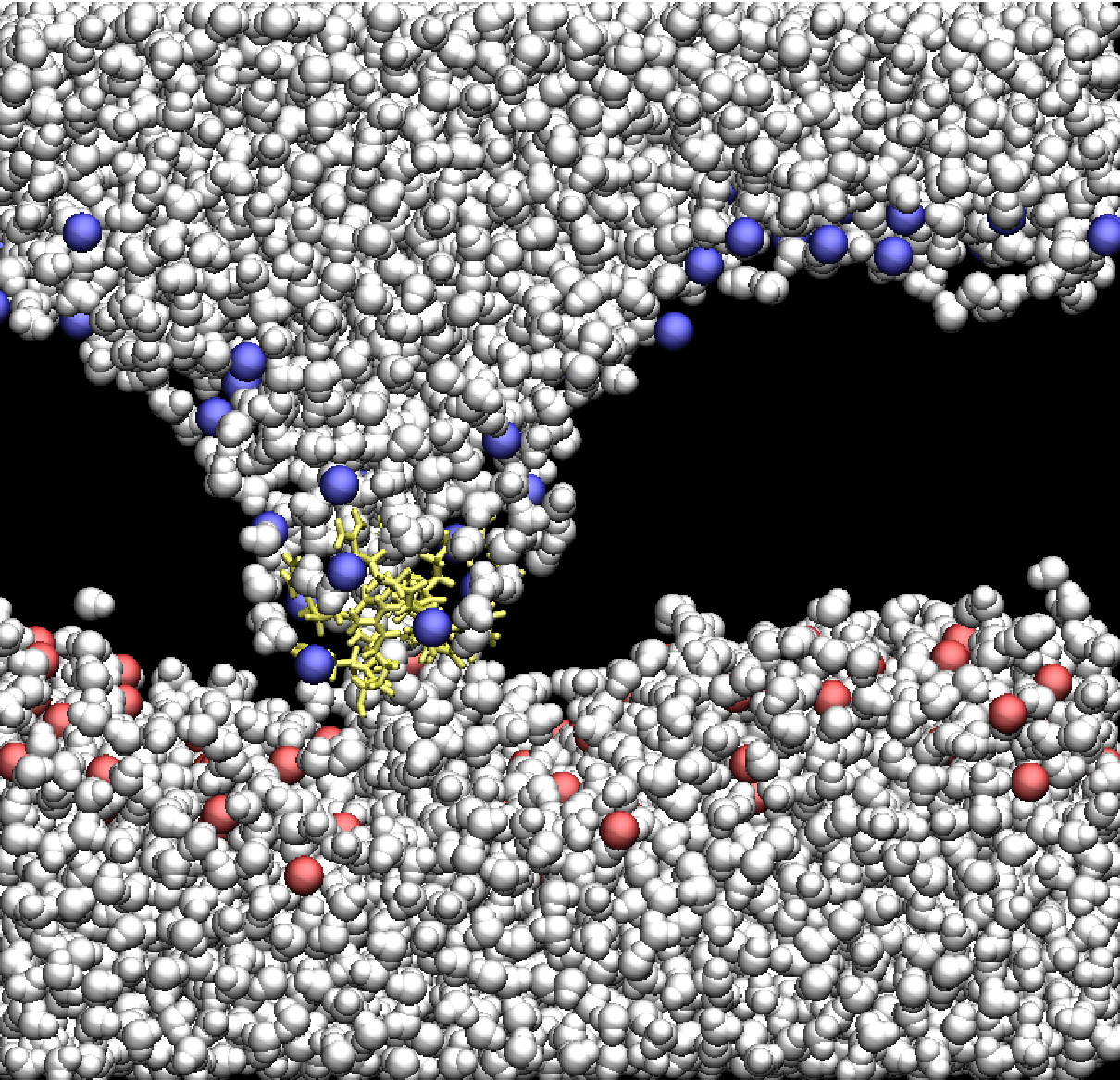}}\vspace{5mm}
\subfloat[the last (25.4 ns)]{\includegraphics[width=4cm,height=4cm]{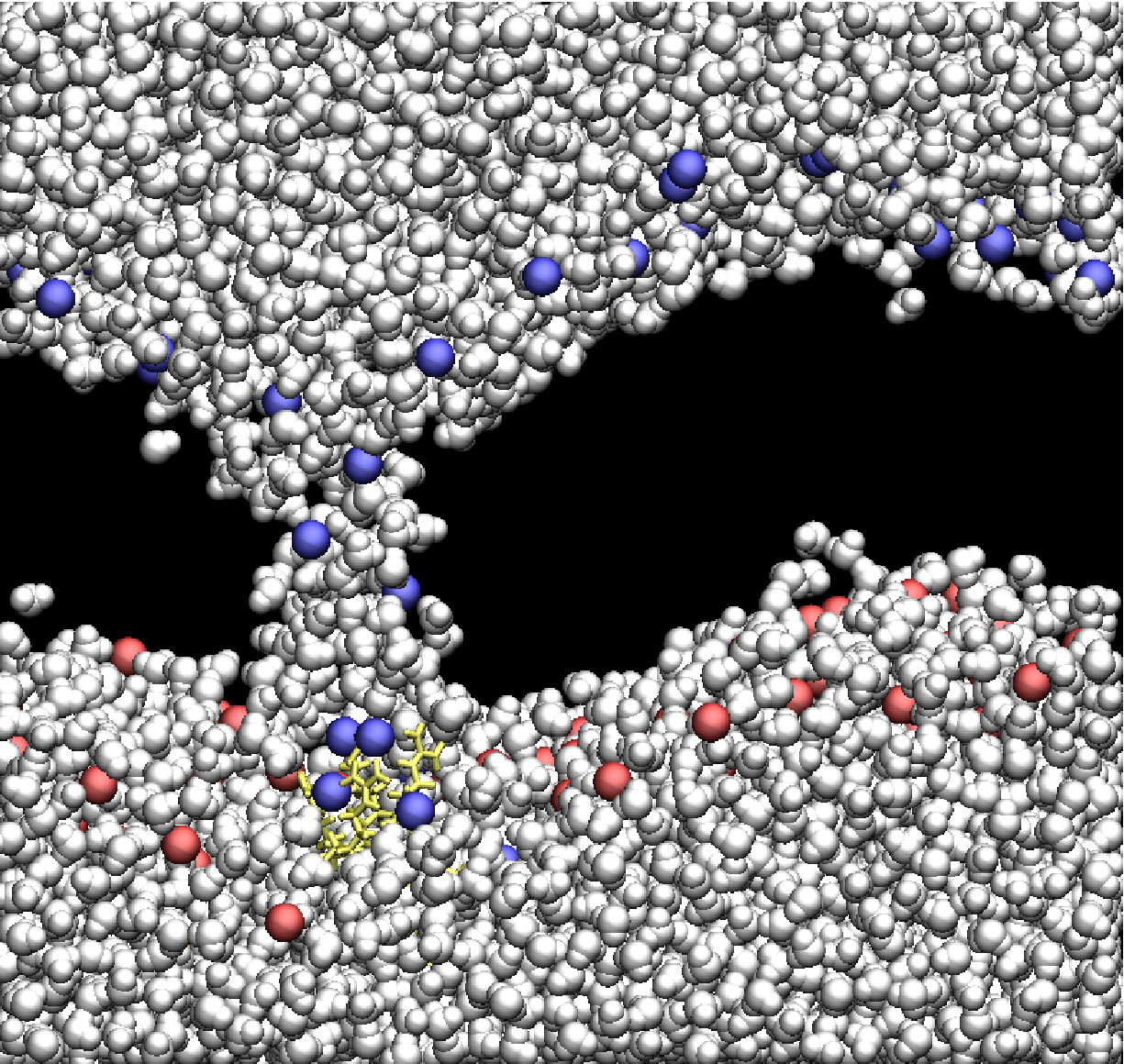}}
\caption{Water molecules near Arg$_9$ at (a) 7.5 ns, (b) 15.0 ns, (c) 22.5 ns, and (d) the last (25.4 ns). (yellow: Arg$_9$, white: water, blue \& red : phosphorus atoms of the upper leaflet \& lower leaflet, respectively). The water flow and the movement of Arg$_9$ are correlated with each other.}\label{fig3}
\end{figure}

\begin{figure}[ht!]%
\centering
\includegraphics[width=8cm,height=6cm]{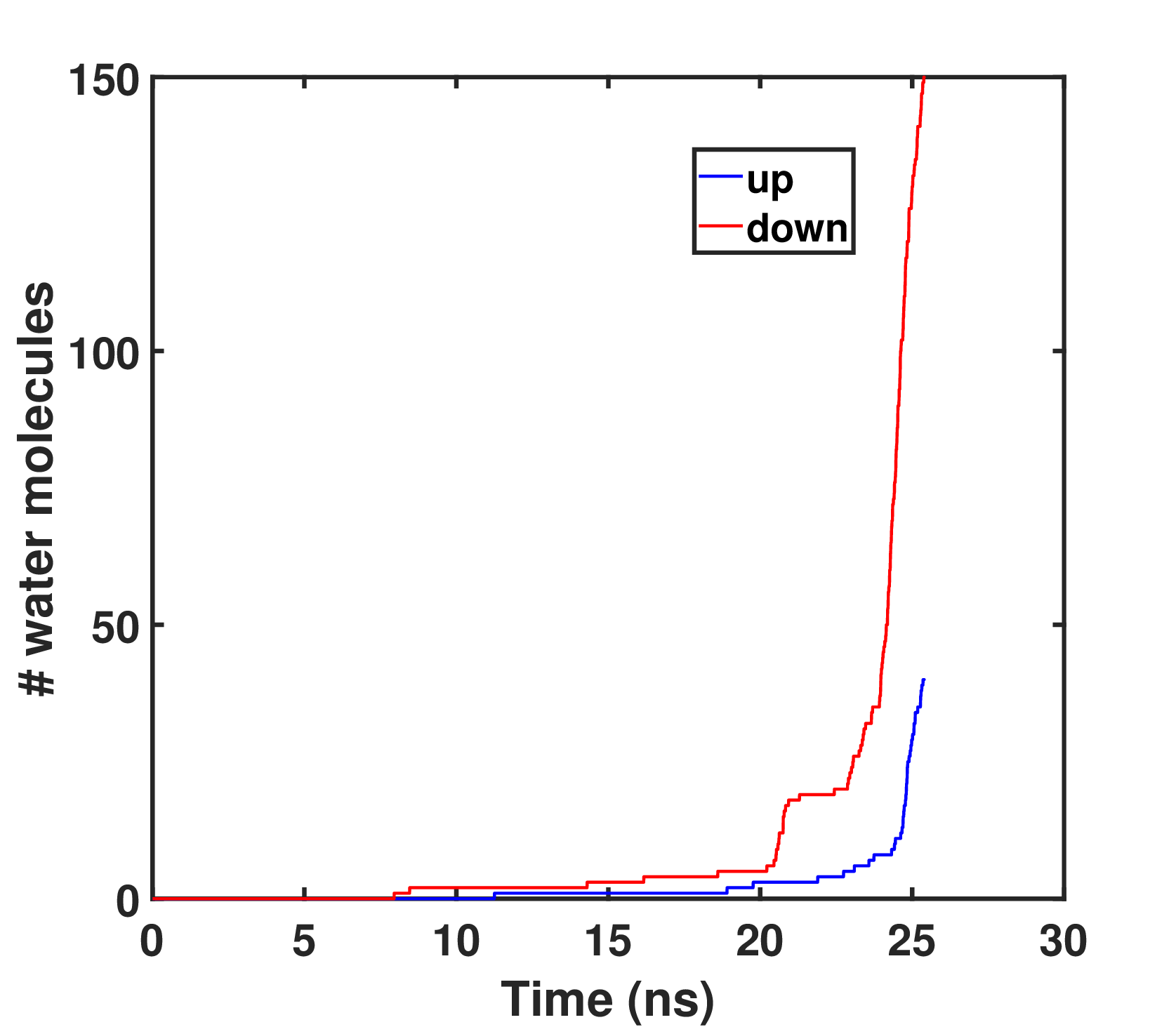}
\caption{The total number of water molecules translocated across the membrane (the moving-up(exiting) \& the moving-down(entering)) vs. the simulation time}\label{fig4}
\end{figure}

When Arg$_9$ approaches the middle of the model membrane, the number of water molecules coordinated with Arg$_9$ increases, as shown in Fig. \ref{fig3}. 
A water pore is shown at 22.5 ns; however, most of the pore is blocked by Arg$_9$. Only a few water molecules can translocate across the membrane. Fig. \ref{fig4} presents the total accumulated number of water molecules translocated across the membrane (up(exiting) \& down(entering)) as a function of time. 

We counted the number of water molecules that translocate across the membrane from z = +10 \AA ~ to z = $-$10 \AA (distances with respect to the membrane center) ~as the moving-down (entering) water, and water molecules from z = $-$10 \AA ~ to z = +10 \AA ~as the moving-up(exiting) water.  It is interesting to see an increase in the moving-down(entering) water molecules when Arg$_9$ approaches the bottom of the membrane. On the other hand, the number of the moving-up(exiting) water molecules slightly changed during the last part of the simulation. Therefore, there is a strong correlation between the direction of water flow and the movement of Arg$_9$. 
The rapid increase in the number of downward(entering) water molecules is closely related to the orientation of Arg$_9$ to the membrane, as shown in the following section.

During the WE simulation, 
the downward(entering) water flux was about 5.9 (water molecules/ns), while the upward(exiting) water flux was 1.6. If we count only the number of water molecules during the last part of translocation after the water pore was made, the downward(entering) water flux was 44.8 (water molecules/ns), and the upward(exiting) water flux was 12.4. 
These numbers can be compared with the previous result (6.2 $\sim$ 27.3), where the water pore was made by moving a single lipid to the center of the bilayer using the umbrella sampling \cite{Bennett2014}. It turns out that the water pore is very transient in our WE simulations, and it is closed soon after Arg$_9$ is translocated to the bottom of the membrane. 
We will discuss more on the closure of the water pore in one of the following sections.

\subsection*{The orientation angle of Arg$_9$ affects the translocation}

As shown in Fig. \ref{fig1}, the penetration depth was increased rapidly during the last part of the translocation (22.5 ns $\sim$). In the previous section, we mentioned that rapid penetration is related to the water flow. In this section, we show that the orientation angle of an Arg$_9$ peptide also affects the efficiency of the translocation. 

\begin{figure}[ht!]%
\centering
\includegraphics[width=8cm,height=6cm]{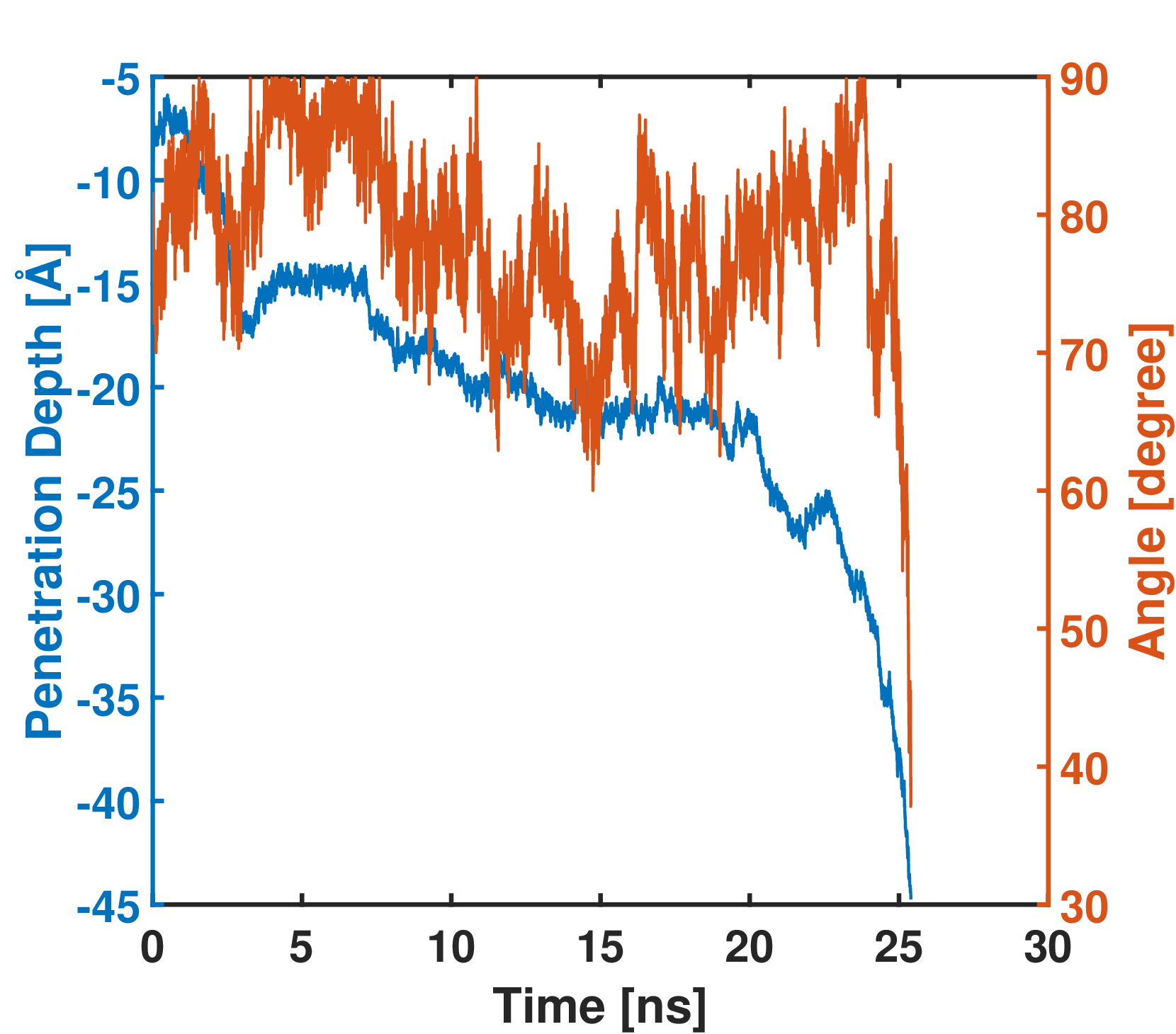}
\caption{The penetration depth (blue line) and the orientation angle (red line) of Arg$_9$ vs. the simulation time.}\label{fig5}
\end{figure}

We define the orientation angle of the peptide as follows: First, obtain the center of mass positions of the first three and the last three C$\alpha$'s of Arg$_9$. Second, find a vector($\vec{v}_1$) connecting those two center-of-mass positions.
Last, calculate an angle between the vector $\vec{v}_1$ and a unit vector parallel to the z-axis (e.g., $\vec{v}_2$ = (0,0,1)). Thus, a smaller angle means that the peptide is almost normal to the model membrane. Note that the conformation of Arg$_9$ is a random coil both in solution and in the membrane environment \cite{Eiriksdottir2010}, and our simulation shows a turn during the translocation (Fig. S2). Therefore, the calculated angle is not as precise
as in the case of a straight helix; however, our calculation will give us an idea of how Arg$_9$ is oriented during the translocation. 
Fig. \ref{fig5} shows the orientation angle of Arg$_9$ during the translocation.
During most simulation time, Arg$_9$ is almost parallel to the surface of the model membrane (70$^\circ$ $\sim$ 90$^\circ$); however, the orientation is rapidly changed to a smaller angle ($\sim$ 40$^\circ$) when the peptide reaches the bottom of the model membrane. Based on this figure, one can conjecture that Arg$_9$ with a smaller orientation angle shows more efficient translocation across the model membrane. This efficiency should be related to free energy barriers between Arg$_9$ and the model membrane.

\begin{figure}[ht!]%
\centering
\includegraphics[width=8cm,height=6cm]{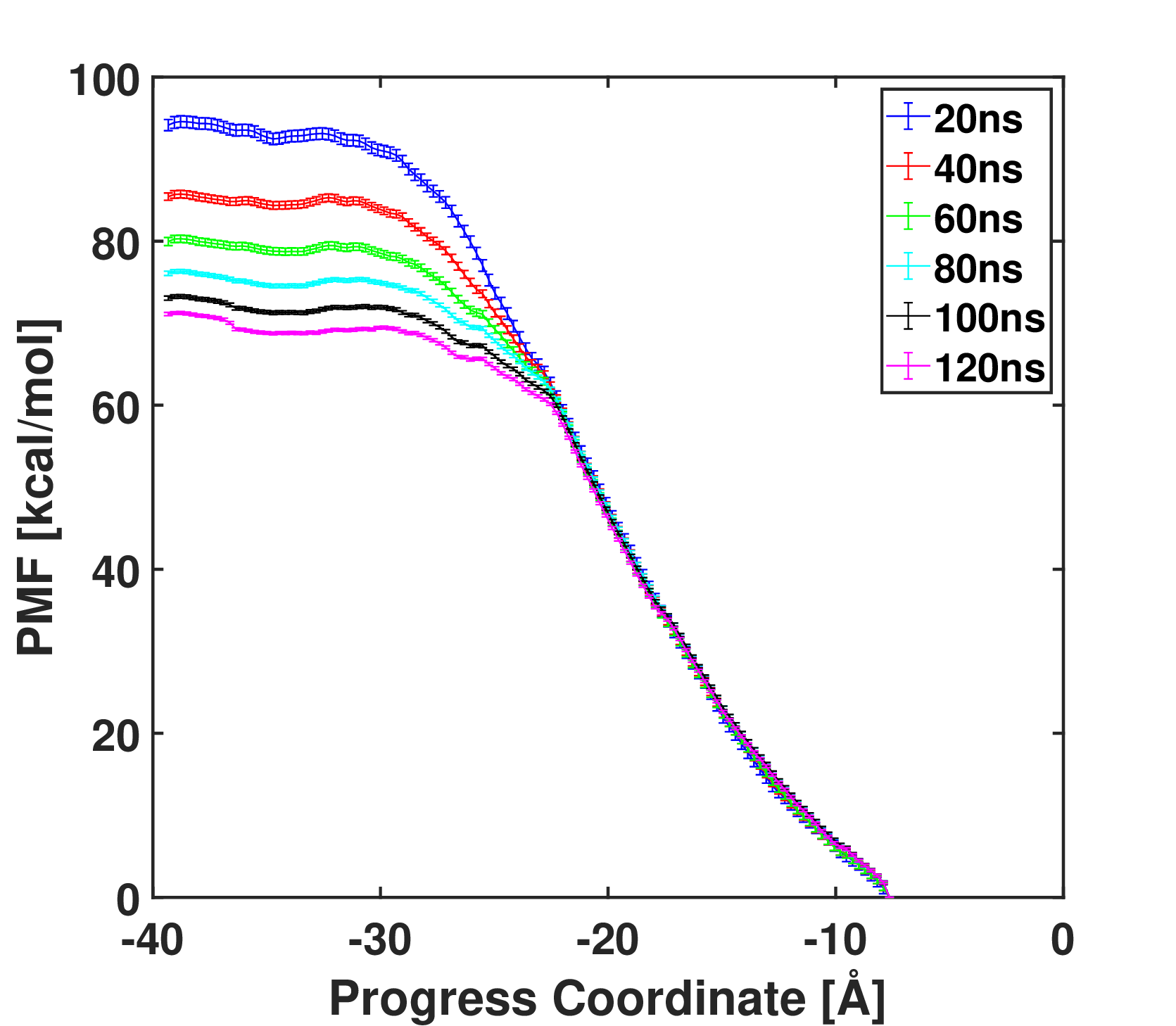}
\caption{The potential of mean force(pmf) vs. the progress coordinate. 20 ns means that we collected the data between 0 to 20 ns (after discarding the first 20 ns data), 40 means 0 to 40 ns data, and so on.}\label{fig6}
\end{figure}

Fig. \ref{fig6} shows the potential of mean force(pmf) along the translocation path of Arg$_9$ using the umbrella sampling. We have 49 windows and sampled 140 ns data at each window. We discarded the first 20 ns data and analyzed the rest of the data (120 ns) for the free energy calculation. In the figure legend, 20 ns means that we collected the data between 0 to 20 ns (after discarding the first 20 ns data), 40 means 0 to 40 ns data, and so on. The pmf plots show convergence, and the error bars in each plot are  getting smaller when increasing the number of data points. The details of the umbrella sampling are given in the Methods section. 
When Arg$_9$ approaches the bottom of the membrane (e.g., the WE4 \& WE5 simulations), the slope of the free energy curve decreases compared with that in the hydrophobic core (e.g., the WE2 \& WE3 simulations). During the WE4 \& WE5 simulations, the penetration depth is drastically increased, and the number of transported water molecules increases. Our simulation confirms the previous simulations \cite{Yesylevskyy2009,Huang2013} that the slope of the potential of mean force becomes smaller in the presence of the water pore. The total free energy barrier from the upper solution to the bottom of the lower leaflet is about 70 kcal/mol. Therefore, the translocation of single Arg$_9$ across the DOPC/DOPG(4:1) membrane is energetically unfavorable even though the water molecules decrease the energy barrier.

Table S1 shows two additional simulations at [$-$45.0 \AA, $-$33.0 \AA], i.e., WE5a and WE5b simulations. We want to compare the translocation efficiency and the orientation angle of Arg$_9$ in each simulation.  In Fig. S3, we compare the penetration depth and the orientation angle from WE5, WE5a, and WE5b simulations.  
The solid blue and red lines correspond to the penetration depth and orientation angle from the WE5 simulation. The results from the WE5a and WE5b simulations were shown as the dashed and dashed-dot lines in the figure. When the penetration depth changes rapidly, the orientation angle shows similar behavior. Whenever Arg$_9$ reaches the bottom of the membrane, the orientation angle becomes smaller (40 $\sim$ 50$^\circ$). The upright position seems efficient for the translocation of Arg$_9$ in our simulation. 
Our simulation suggests that the water flow and the orientation angle of Arg$_9$ play a role in translocation. We observed only a single translocation; more translocated trajectories are needed to confirm our findings. Furthermore, we might need other observables to quantify the orientation of Arg$_9$ during the translocation.  

\subsection*{The water pore was very transient and lipid flip-flops were observed}

When Arg$_9$ was located at the bottom of the membrane, the WE5 simulation was stopped, and 
an additional conventional MD simulation (without using the WE method) was run at this position. The simulation method is the same as in the Methods section and in the previous work \cite{Choe2020,Choe2021}.
A total of 1 $\mu$s simulation was performed. During the simulation, Arg$_9$ moved up and down for a while, and eventually, it moved to the bottom of the membrane and stayed there until the end of the simulation. 
The water pore was closed at 70 ns, and there was no translocation of water molecules after that. 
Fig. S4 presents several snapshots during this simulation which show a closure of the water pore and the behavior of Arg$_9$.
The pore lifetime in our simulations (WE simulations + an additional conventional simulation) is less than 100 ns (Fig. S5), and the order of lifetime is very similar to previous results for other systems \cite{Marrink2001,deVries2004}.

Another interesting finding during the simulation was the translocation of lipids molecules. 
The phospholipids(PL) flip-flop (or PL translocation) has been well studied in numerous works \cite{Fattal1994,Matsuzaki1996,Kol2003,Tieleman2006,Gurtovenko2007,Sun2014,Parosio2016,Allhusen2017,Nguyen2021}. There are two primary mechanisms for the lipid flip-flop: One is water pore-mediated. The other one is peptide-mediated.
As shown in Fig. \ref{fig2}, a few lipids in the upper leaflet moved down along with Arg$_9$. A total of six DOPG lipids
moved down to the lower leaflet during the WE simulation, and two of them moved back to the upper leaflet when Arg$_9$ moved up during the conventional simulation (Fig. S4). Our simulation shows that the lipids' location was switched rapidly from the upper leaflet to the lower leaflet and vice versa. The total time of PL translocation (moved from the upper leaflet to the lower and then moved back to the upper leaflet) was about 75 ns. This time scale is due to the fast movement of Arg$_9$ in our simulations.
Our simulations suggest that the primary mechanism of the lipid translocation is a peptide-mediated process; however, the water flow also affects PL translocation because the peptide movement and the water flow are closely related, as shown in the WE simulations and the additional conventional simulation. 
We found that the translocated lipids were quickly reorganized (Fig. S4). After the translocation of lipids to the lower or the upper leaflet, the flip-flopped lipids were well-reorganized with the other lipids during the conventional simulation.

\section*{Discussion}\label{sec4}

Translocation of a single Arg$_9$ was observed using the WE method within the MD simulations approach. 
Compared with the previous WE simulation \cite{Choe2021}, which didn't show any translocation of Arg$_9$ across the membrane, the smaller bin size is one of the critical factors that made it possible to observe a spontaneous translocation across the model membrane. Finding an appropriate number of walkers(replicas) and a reasonable bin size in the WE simulations is difficult without trial and error. Note that an adaptive bin scheme was already developed for the WE simulation to change the bin size on-the-fly \cite{Torrillo2021}. This scheme could be applied to the current work to make rapid translocation of Arg$_9$ or any other CPPs to study their transport mechanisms.    
Our WE simulations showed that the translocation of a single Arg$_9$ was energetically unfavorable due to higher free energy barriers ($\sim$ 70 kcal/mol) between Arg$_9$ and the model membrane along the translocation path. 

It has been known that one of the essential factors in the translocation of CPPs is the concentration of peptides. Although we used four Arg$_9$s in our WE simulation, the concentration in a local area is insufficient to see collective behavior between Arg$_9$s. 
The progress coordinate in our WE simulation was defined as the distance between the upper leaflet and one of the Arg$_9$s, and thus the other three Arg$_9$s didn't show any penetration into the model membrane during the whole simulation.
The collective behavior between CPPs 
could enhance the efficiency of translocation. Although our simulation shows that the translocation of a single CPP is not favorable, the WE method can be used to study spontaneous translocation at low concentrations. One can
implement a progress coordinate as a distance between the center of mass of the upper leaflet's phosphorus atoms and more than two CPPs, e.g., two Arg$_9$s, three Arg$_9$s, or four Arg$_9$s at the same time. However, our experience in the WE simulations shows that it is not easy to observe  simultaneous translocation of more than two CPPs due to much slower CPPs than a single CPP movement.  

Another critical factor in our successful translocation is the orientation of Arg$_9$ relative to the membrane. 
Our simulation showed that the orientation angle of Arg$_9$ is essential to determining the energy barrier along the translocation path of Arg$_9$.  
It is much easier for Arg$_9$, which was oriented along the normal of the membrane surface (i.e., parallel to the z-axis), to translocate across the membrane, as shown in Fig. S3. It has been known that the peptides' orientation depends on the concentration. For example, a two-state model has been suggested to orient adsorbed peptides: S-state \& I-state \cite{Huang2000}.
Here, the S-state means that peptides are parallel to the membrane and observed at low P/Ls, while the I-state denotes that peptides are perpendicular to the membrane and observed above the threshold concentration. 
To stabilize the water pore, it is necessary to change the average peptide orientation from S to I \cite{Huang2000,Sun2014}. 
The threshold value (P/L$^{*}$)
is the peptide concentration 
when the energy levels of the S and I states are equal.  
Although Huang's two-state model was initially developed for  antimicrobial peptides, 
the orientation of Arg$_9$ relative to the membrane in our simulations can also depend on the P/L ratio. The P/L ratio in our system is low (P/L $\sim$ 0.042); however, this ratio is much greater than the critical thresholds in the literature \cite{Huang2004,Sun2014}.
Note that the threshold value (P/L$^{*}$) varies with the lipid composition of the membrane \cite{Huang2000,Huang2004,Sun2014}.
Arg$_9$ stayed at the S state during most of the simulation time and turned into the I state during the last part of the simulation. Although we found the translocation of Arg$_9$ across the model membrane within a very short time (a few tens ns), it is necessary to work on different lipid compositions and concentrations to see if there is any significant change in this time scale. 

Although we observed a spontaneous translocation of a single Arg$_9$ within a very short time scale, our results have some limitations. First, there is a considerable hysteresis in the PMF profile using the umbrella sampling due to asymmetric membrane distortions, as shown in Fig. \ref{fig2}. The opening of the water pore did not happen until the peptide reached the lower leaflet, and this makes the PMF plot asymmetric around the center of the membrane. We think that the orientation angle of Arg$_9$ also contributes to the hysteresis in the PMF plot. A total of 140 ns of sampling in each window may not be enough to sample equilibrium configurations. Second, the external electric field was applied in all the simulations, including the WE simulations, and this electric field could contribute to the significant distortions of the membrane because our membrane has asymmetric distributions of DOPG lipids in both layers.  It has been shown that the penetration of R8 within  coarse-grained simulations depends on the external electric field \cite{Wang2018}, and the authors showed that a higher electric field resulted in a short penetration time. The free energy calculations were also dependent on the electrostatic properties of the membrane \cite{Via2018}. Although our electric field strength (0.05 V/nm) is smaller than Wang et al.'s (0.17 $\sim$ 0.20 V/nm) \cite{Wang2018}, we can not neglect the electric field effect on our charged Arg$_9$ and its translocation across the membrane.  We may need another WE simulation without the external electric field to see the changes in the translocation of charged peptides across the charged membrane. Third, the translocation time scale reported here can not be compared with experiments. In our current simulation setup, it is challenging to compute transition probabilities between bins and thus a mean first passage time. Short time scales in our current work are only computer simulation time. Last, although we observed the translocation of Arg$_9$ within a very short time scale, the umbrella sampling showed that the free energy barrier is still large for a single Arg$_9$ to translocate across the membrane. The single reaction coordinate (the penetration depth in the z-direction) in our umbrella sampling may not be enough to identify the actual translocation paths of Arg$_9$ and the free energy barriers along the paths. One can add an additional reaction coordinate (e.g., the orientation angle of Arg$_9$) in the umbrella sampling to see if there is  any change in the free energy barrier.   

Our WE simulations suggest that the WE method can effectively sample rare events, such as a spontaneous translocation of CPP within MD simulations approach. Furthermore, we expect the WE method to be applied to any CPPs to reveal interactions between CPPs and various membranes. Moreover, the WE method can provide insights into the transport mechanisms of various membrane-active peptides(MAPs).

\bibliography{cpp}

\begin{thebibliography}{54}%
\makeatletter
\providecommand \@ifxundefined [1]{%
 \@ifx{#1\undefined}
}%
\providecommand \@ifnum [1]{%
 \ifnum #1\expandafter \@firstoftwo
 \else \expandafter \@secondoftwo
 \fi
}%
\providecommand \@ifx [1]{%
 \ifx #1\expandafter \@firstoftwo
 \else \expandafter \@secondoftwo
 \fi
}%
\providecommand \natexlab [1]{#1}%
\providecommand \enquote  [1]{``#1''}%
\providecommand \bibnamefont  [1]{#1}%
\providecommand \bibfnamefont [1]{#1}%
\providecommand \citenamefont [1]{#1}%
\providecommand \href@noop [0]{\@secondoftwo}%
\providecommand \href [0]{\begingroup \@sanitize@url \@href}%
\providecommand \@href[1]{\@@startlink{#1}\@@href}%
\providecommand \@@href[1]{\endgroup#1\@@endlink}%
\providecommand \@sanitize@url [0]{\catcode `\\12\catcode `\$12\catcode
  `\&12\catcode `\#12\catcode `\^12\catcode `\_12\catcode `\%12\relax}%
\providecommand \@@startlink[1]{}%
\providecommand \@@endlink[0]{}%
\providecommand \url  [0]{\begingroup\@sanitize@url \@url }%
\providecommand \@url [1]{\endgroup\@href {#1}{\urlprefix }}%
\providecommand \urlprefix  [0]{URL }%
\providecommand \Eprint [0]{\href }%
\providecommand \doibase [0]{https://doi.org/}%
\providecommand \selectlanguage [0]{\@gobble}%
\providecommand \bibinfo  [0]{\@secondoftwo}%
\providecommand \bibfield  [0]{\@secondoftwo}%
\providecommand \translation [1]{[#1]}%
\providecommand \BibitemOpen [0]{}%
\providecommand \bibitemStop [0]{}%
\providecommand \bibitemNoStop [0]{.\EOS\space}%
\providecommand \EOS [0]{\spacefactor3000\relax}%
\providecommand \BibitemShut  [1]{\csname bibitem#1\endcsname}%
\let\auto@bib@innerbib\@empty
\bibitem [{\citenamefont {Guidotti}\ \emph {et~al.}(2017)\citenamefont
  {Guidotti}, \citenamefont {Brambilla},\ and\ \citenamefont
  {Rossi}}]{Guidotti2017}%
  \BibitemOpen
  \bibfield  {author} {\bibinfo {author} {\bibfnamefont {G.}~\bibnamefont
  {Guidotti}}, \bibinfo {author} {\bibfnamefont {L.}~\bibnamefont
  {Brambilla}},\ and\ \bibinfo {author} {\bibfnamefont {D.}~\bibnamefont
  {Rossi}},\ }\href@noop {} {\bibfield  {journal} {\bibinfo  {journal} {Trends
  in Pharmacological Sciences}\ }\textbf {\bibinfo {volume} {38}},\ \bibinfo
  {pages} {406} (\bibinfo {year} {2017})}\BibitemShut {NoStop}%
\bibitem [{\citenamefont {Walrant}\ \emph {et~al.}(2017)\citenamefont
  {Walrant}, \citenamefont {Cardon}, \citenamefont {Burlina},\ and\
  \citenamefont {Sagan}}]{Walrant2017}%
  \BibitemOpen
  \bibfield  {author} {\bibinfo {author} {\bibfnamefont {A.}~\bibnamefont
  {Walrant}}, \bibinfo {author} {\bibfnamefont {S.}~\bibnamefont {Cardon}},
  \bibinfo {author} {\bibfnamefont {F.}~\bibnamefont {Burlina}},\ and\ \bibinfo
  {author} {\bibfnamefont {S.}~\bibnamefont {Sagan}},\ }\href@noop {}
  {\bibfield  {journal} {\bibinfo  {journal} {Accounts Chem Res.}\ }\textbf
  {\bibinfo {volume} {50}},\ \bibinfo {pages} {2968–2975} (\bibinfo {year}
  {2017})}\BibitemShut {NoStop}%
\bibitem [{\citenamefont {Pisa}\ \emph {et~al.}(2015)\citenamefont {Pisa},
  \citenamefont {Chassaing},\ and\ \citenamefont {Swiecicki}}]{Pisa2014}%
  \BibitemOpen
  \bibfield  {author} {\bibinfo {author} {\bibfnamefont {M.~D.}\ \bibnamefont
  {Pisa}}, \bibinfo {author} {\bibfnamefont {G.}~\bibnamefont {Chassaing}},\
  and\ \bibinfo {author} {\bibfnamefont {J.}~\bibnamefont {Swiecicki}},\
  }\href@noop {} {\bibfield  {journal} {\bibinfo  {journal} {Biochemistry}\
  }\textbf {\bibinfo {volume} {54}},\ \bibinfo {pages} {194} (\bibinfo {year}
  {2015})}\BibitemShut {NoStop}%
\bibitem [{\citenamefont {Ruseska}\ and\ \citenamefont
  {Zimmer}(2020)}]{Ruseska2020}%
  \BibitemOpen
  \bibfield  {author} {\bibinfo {author} {\bibfnamefont {I.}~\bibnamefont
  {Ruseska}}\ and\ \bibinfo {author} {\bibfnamefont {A.}~\bibnamefont
  {Zimmer}},\ }\href@noop {} {\bibfield  {journal} {\bibinfo  {journal}
  {Beilstein J. of Nanotechnol.}\ }\textbf {\bibinfo {volume} {11}},\ \bibinfo
  {pages} {101} (\bibinfo {year} {2020})}\BibitemShut {NoStop}%
\bibitem [{\citenamefont {Madani}\ \emph {et~al.}(2011)\citenamefont {Madani},
  \citenamefont {Lindberg}, \citenamefont {Langel}, \citenamefont {Futaki},\
  and\ \citenamefont {Gr\"{a}slund}}]{Madani2011}%
  \BibitemOpen
  \bibfield  {author} {\bibinfo {author} {\bibfnamefont {F.}~\bibnamefont
  {Madani}}, \bibinfo {author} {\bibfnamefont {S.}~\bibnamefont {Lindberg}},
  \bibinfo {author} {\bibfnamefont {U.}~\bibnamefont {Langel}}, \bibinfo
  {author} {\bibfnamefont {S.}~\bibnamefont {Futaki}},\ and\ \bibinfo {author}
  {\bibfnamefont {A.}~\bibnamefont {Gr\"{a}slund}},\ }\href@noop {} {\bibfield
  {journal} {\bibinfo  {journal} {J. Biophys.}\ }\textbf {\bibinfo {volume}
  {2011}},\ \bibinfo {pages} {1} (\bibinfo {year} {2011})}\BibitemShut
  {NoStop}%
\bibitem [{\citenamefont {Herce}\ and\ \citenamefont
  {Garcia}(2007)}]{Herce2007}%
  \BibitemOpen
  \bibfield  {author} {\bibinfo {author} {\bibfnamefont {H.~D.}\ \bibnamefont
  {Herce}}\ and\ \bibinfo {author} {\bibfnamefont {A.~E.}\ \bibnamefont
  {Garcia}},\ }\href@noop {} {\bibfield  {journal} {\bibinfo  {journal} {Proc.
  Natl. Acad. Sci. USA}\ }\textbf {\bibinfo {volume} {104}},\ \bibinfo {pages}
  {20805} (\bibinfo {year} {2007})}\BibitemShut {NoStop}%
\bibitem [{\citenamefont {Herce}\ \emph {et~al.}(2009)\citenamefont {Herce},
  \citenamefont {Garcia}, \citenamefont {Litt}, \citenamefont {Kane},
  \citenamefont {Martin}, \citenamefont {Enrique}, \citenamefont {Rebolledo},\
  and\ \citenamefont {Milesi}}]{Herce2009}%
  \BibitemOpen
  \bibfield  {author} {\bibinfo {author} {\bibfnamefont {H.~D.}\ \bibnamefont
  {Herce}}, \bibinfo {author} {\bibfnamefont {A.~E.}\ \bibnamefont {Garcia}},
  \bibinfo {author} {\bibfnamefont {J.}~\bibnamefont {Litt}}, \bibinfo {author}
  {\bibfnamefont {R.~S.}\ \bibnamefont {Kane}}, \bibinfo {author}
  {\bibfnamefont {P.}~\bibnamefont {Martin}}, \bibinfo {author} {\bibfnamefont
  {N.}~\bibnamefont {Enrique}}, \bibinfo {author} {\bibfnamefont
  {A.}~\bibnamefont {Rebolledo}},\ and\ \bibinfo {author} {\bibfnamefont
  {V.}~\bibnamefont {Milesi}},\ }\href@noop {} {\bibfield  {journal} {\bibinfo
  {journal} {Biophys. J.}\ }\textbf {\bibinfo {volume} {97}},\ \bibinfo {pages}
  {1917} (\bibinfo {year} {2009})}\BibitemShut {NoStop}%
\bibitem [{\citenamefont {Yesylevskyy}\ \emph {et~al.}(2009)\citenamefont
  {Yesylevskyy}, \citenamefont {Marrink},\ and\ \citenamefont
  {Mark}}]{Yesylevskyy2009}%
  \BibitemOpen
  \bibfield  {author} {\bibinfo {author} {\bibfnamefont {S.}~\bibnamefont
  {Yesylevskyy}}, \bibinfo {author} {\bibfnamefont {S.-J.}\ \bibnamefont
  {Marrink}},\ and\ \bibinfo {author} {\bibfnamefont {A.}~\bibnamefont
  {Mark}},\ }\href@noop {} {\bibfield  {journal} {\bibinfo  {journal} {Biophys.
  J.}\ }\textbf {\bibinfo {volume} {97}},\ \bibinfo {pages} {40} (\bibinfo
  {year} {2009})}\BibitemShut {NoStop}%
\bibitem [{\citenamefont {Zorko}\ and\ \citenamefont
  {Langel}(2005)}]{Zorko2005}%
  \BibitemOpen
  \bibfield  {author} {\bibinfo {author} {\bibfnamefont {M.}~\bibnamefont
  {Zorko}}\ and\ \bibinfo {author} {\bibfnamefont {{\"U}.}~\bibnamefont
  {Langel}},\ }\href@noop {} {\bibfield  {journal} {\bibinfo  {journal} {Adv.
  Drug Deliv. Rev.}\ }\textbf {\bibinfo {volume} {57}},\ \bibinfo {pages} {529}
  (\bibinfo {year} {2005})}\BibitemShut {NoStop}%
\bibitem [{\citenamefont {Pourmousa}\ \emph {et~al.}(2013)\citenamefont
  {Pourmousa}, \citenamefont {Wong-ekkabut}, \citenamefont {Patra},\ and\
  \citenamefont {Karttunen}}]{Pourmousa2013}%
  \BibitemOpen
  \bibfield  {author} {\bibinfo {author} {\bibfnamefont {M.}~\bibnamefont
  {Pourmousa}}, \bibinfo {author} {\bibfnamefont {J.}~\bibnamefont
  {Wong-ekkabut}}, \bibinfo {author} {\bibfnamefont {M.}~\bibnamefont
  {Patra}},\ and\ \bibinfo {author} {\bibfnamefont {M.}~\bibnamefont
  {Karttunen}},\ }\href@noop {} {\bibfield  {journal} {\bibinfo  {journal} {J.
  Phys. Chem. B}\ }\textbf {\bibinfo {volume} {117}},\ \bibinfo {pages} {230}
  (\bibinfo {year} {2013})}\BibitemShut {NoStop}%
\bibitem [{\citenamefont {Sun}\ \emph {et~al.}(2014)\citenamefont {Sun},
  \citenamefont {Forsman}, \citenamefont {Lund},\ and\ \citenamefont
  {Woodward}}]{Sun2014}%
  \BibitemOpen
  \bibfield  {author} {\bibinfo {author} {\bibfnamefont {D.}~\bibnamefont
  {Sun}}, \bibinfo {author} {\bibfnamefont {J.}~\bibnamefont {Forsman}},
  \bibinfo {author} {\bibfnamefont {M.}~\bibnamefont {Lund}},\ and\ \bibinfo
  {author} {\bibfnamefont {C.~E.}\ \bibnamefont {Woodward}},\ }\href@noop {}
  {\bibfield  {journal} {\bibinfo  {journal} {Phys.Chem.Chem.Phys.}\ }\textbf
  {\bibinfo {volume} {16}},\ \bibinfo {pages} {20785} (\bibinfo {year}
  {2014})}\BibitemShut {NoStop}%
\bibitem [{\citenamefont {Teseia}\ \emph {et~al.}(2017)\citenamefont {Teseia},
  \citenamefont {Vazdarb}, \citenamefont {Jensenc}, \citenamefont {Cragnella},
  \citenamefont {Masond}, \citenamefont {Heydae}, \citenamefont {Skepo},
  \citenamefont {Jungwirthd},\ and\ \citenamefont {Lunda}}]{Tesei2017}%
  \BibitemOpen
  \bibfield  {author} {\bibinfo {author} {\bibfnamefont {G.}~\bibnamefont
  {Teseia}}, \bibinfo {author} {\bibfnamefont {M.}~\bibnamefont {Vazdarb}},
  \bibinfo {author} {\bibfnamefont {M.~R.}\ \bibnamefont {Jensenc}}, \bibinfo
  {author} {\bibfnamefont {C.}~\bibnamefont {Cragnella}}, \bibinfo {author}
  {\bibfnamefont {P.~E.}\ \bibnamefont {Masond}}, \bibinfo {author}
  {\bibfnamefont {J.}~\bibnamefont {Heydae}}, \bibinfo {author} {\bibfnamefont
  {M.}~\bibnamefont {Skepo}}, \bibinfo {author} {\bibfnamefont
  {P.}~\bibnamefont {Jungwirthd}},\ and\ \bibinfo {author} {\bibfnamefont
  {M.}~\bibnamefont {Lunda}},\ }\href@noop {} {\bibfield  {journal} {\bibinfo
  {journal} {PNAS}\ }\textbf {\bibinfo {volume} {114}},\ \bibinfo {pages}
  {11428} (\bibinfo {year} {2017})}\BibitemShut {NoStop}%
\bibitem [{\citenamefont {Yao}\ \emph {et~al.}(2019)\citenamefont {Yao},
  \citenamefont {Kang}, \citenamefont {Yu}, \citenamefont {Chen}, \citenamefont
  {Liu},\ and\ \citenamefont {Wang}}]{Yao2019}%
  \BibitemOpen
  \bibfield  {author} {\bibinfo {author} {\bibfnamefont {C.}~\bibnamefont
  {Yao}}, \bibinfo {author} {\bibfnamefont {Z.}~\bibnamefont {Kang}}, \bibinfo
  {author} {\bibfnamefont {B.}~\bibnamefont {Yu}}, \bibinfo {author}
  {\bibfnamefont {Q.}~\bibnamefont {Chen}}, \bibinfo {author} {\bibfnamefont
  {Y.}~\bibnamefont {Liu}},\ and\ \bibinfo {author} {\bibfnamefont
  {Q.}~\bibnamefont {Wang}},\ }\href@noop {} {\bibfield  {journal} {\bibinfo
  {journal} {Langmuir}\ }\textbf {\bibinfo {volume} {35}},\ \bibinfo {pages}
  {9286–9296} (\bibinfo {year} {2019})}\BibitemShut {NoStop}%
\bibitem [{\citenamefont {Choong}\ and\ \citenamefont
  {Yap}(2021)}]{Choong2021}%
  \BibitemOpen
  \bibfield  {author} {\bibinfo {author} {\bibfnamefont {F.~H.}\ \bibnamefont
  {Choong}}\ and\ \bibinfo {author} {\bibfnamefont {B.~K.}\ \bibnamefont
  {Yap}},\ }\href@noop {} {\bibfield  {journal} {\bibinfo  {journal}
  {ChemPhysChem}\ }\textbf {\bibinfo {volume} {22}},\ \bibinfo {pages} {493}
  (\bibinfo {year} {2021})}\BibitemShut {NoStop}%
\bibitem [{\citenamefont {Akhunzada}\ \emph {et~al.}(2017)\citenamefont
  {Akhunzada}, \citenamefont {Chandramouli}, \citenamefont {Bhattacharjee},
  \citenamefont {Macchi}, \citenamefont {Cardarelli},\ and\ \citenamefont
  {Brancato}}]{Akhunzada2017}%
  \BibitemOpen
  \bibfield  {author} {\bibinfo {author} {\bibfnamefont {M.~J.}\ \bibnamefont
  {Akhunzada}}, \bibinfo {author} {\bibfnamefont {B.}~\bibnamefont
  {Chandramouli}}, \bibinfo {author} {\bibfnamefont {N.}~\bibnamefont
  {Bhattacharjee}}, \bibinfo {author} {\bibfnamefont {S.}~\bibnamefont
  {Macchi}}, \bibinfo {author} {\bibfnamefont {F.}~\bibnamefont {Cardarelli}},\
  and\ \bibinfo {author} {\bibfnamefont {G.}~\bibnamefont {Brancato}},\
  }\href@noop {} {\bibfield  {journal} {\bibinfo  {journal} {Phys. Chem. Chem.
  Phys.}\ }\textbf {\bibinfo {volume} {19}},\ \bibinfo {pages} {27603}
  (\bibinfo {year} {2017})}\BibitemShut {NoStop}%
\bibitem [{\citenamefont {Fretz}\ \emph {et~al.}(2007)\citenamefont {Fretz},
  \citenamefont {Penning}, \citenamefont {Al-Taei}, \citenamefont {Futaki},
  \citenamefont {Takeuchi}, \citenamefont {Nakase}, \citenamefont {Storm},\
  and\ \citenamefont {Jones}}]{Fretz07}%
  \BibitemOpen
  \bibfield  {author} {\bibinfo {author} {\bibfnamefont {M.~M.}\ \bibnamefont
  {Fretz}}, \bibinfo {author} {\bibfnamefont {N.~A.}\ \bibnamefont {Penning}},
  \bibinfo {author} {\bibfnamefont {S.}~\bibnamefont {Al-Taei}}, \bibinfo
  {author} {\bibfnamefont {S.}~\bibnamefont {Futaki}}, \bibinfo {author}
  {\bibfnamefont {T.}~\bibnamefont {Takeuchi}}, \bibinfo {author}
  {\bibfnamefont {I.}~\bibnamefont {Nakase}}, \bibinfo {author} {\bibfnamefont
  {G.}~\bibnamefont {Storm}},\ and\ \bibinfo {author} {\bibfnamefont {A.~T.}\
  \bibnamefont {Jones}},\ }\href@noop {} {\bibfield  {journal} {\bibinfo
  {journal} {Biochem. J.}\ }\textbf {\bibinfo {volume} {403}},\ \bibinfo
  {pages} {335} (\bibinfo {year} {2007})}\BibitemShut {NoStop}%
\bibitem [{\citenamefont {Huber}\ and\ \citenamefont {Kim}(1996)}]{Huber1996}%
  \BibitemOpen
  \bibfield  {author} {\bibinfo {author} {\bibfnamefont {G.}~\bibnamefont
  {Huber}}\ and\ \bibinfo {author} {\bibfnamefont {S.}~\bibnamefont {Kim}},\
  }\href@noop {} {\bibfield  {journal} {\bibinfo  {journal} {Biophys J.}\
  }\textbf {\bibinfo {volume} {70}},\ \bibinfo {pages} {97–110} (\bibinfo
  {year} {1996})}\BibitemShut {NoStop}%
\bibitem [{\citenamefont {Zuckerman}\ and\ \citenamefont
  {Chong}(2017)}]{Zuckerman2017}%
  \BibitemOpen
  \bibfield  {author} {\bibinfo {author} {\bibfnamefont {D.~M.}\ \bibnamefont
  {Zuckerman}}\ and\ \bibinfo {author} {\bibfnamefont {L.~T.}\ \bibnamefont
  {Chong}},\ }\href@noop {} {\bibfield  {journal} {\bibinfo  {journal} {Annu.
  Rev. Biophys.}\ }\textbf {\bibinfo {volume} {46}},\ \bibinfo {pages} {43}
  (\bibinfo {year} {2017})}\BibitemShut {NoStop}%
\bibitem [{\citenamefont {Choe}(2021)}]{Choe2021}%
  \BibitemOpen
  \bibfield  {author} {\bibinfo {author} {\bibfnamefont {S.}~\bibnamefont
  {Choe}},\ }\href@noop {} {\bibfield  {journal} {\bibinfo  {journal}
  {Membranes}\ }\textbf {\bibinfo {volume} {11}},\ \bibinfo {pages} {974}
  (\bibinfo {year} {2021})}\BibitemShut {NoStop}%
\bibitem [{\citenamefont {Zwier}\ \emph {et~al.}(2015)\citenamefont {Zwier},
  \citenamefont {Adelman}, \citenamefont {Kaus}, \citenamefont {Pratt},
  \citenamefont {Wong}, \citenamefont {Rego}, \citenamefont {Suarez},
  \citenamefont {Lettieri}, \citenamefont {Wang}, \citenamefont {Grabe},
  \citenamefont {Zuckerman},\ and\ \citenamefont {Chong}}]{westpa}%
  \BibitemOpen
  \bibfield  {author} {\bibinfo {author} {\bibfnamefont {M.~C.}\ \bibnamefont
  {Zwier}}, \bibinfo {author} {\bibfnamefont {J.~L.}\ \bibnamefont {Adelman}},
  \bibinfo {author} {\bibfnamefont {J.~W.}\ \bibnamefont {Kaus}}, \bibinfo
  {author} {\bibfnamefont {A.~J.}\ \bibnamefont {Pratt}}, \bibinfo {author}
  {\bibfnamefont {K.~F.}\ \bibnamefont {Wong}}, \bibinfo {author}
  {\bibfnamefont {N.~B.}\ \bibnamefont {Rego}}, \bibinfo {author}
  {\bibfnamefont {E.}~\bibnamefont {Suarez}}, \bibinfo {author} {\bibfnamefont
  {S.}~\bibnamefont {Lettieri}}, \bibinfo {author} {\bibfnamefont {D.~W.}\
  \bibnamefont {Wang}}, \bibinfo {author} {\bibfnamefont {M.}~\bibnamefont
  {Grabe}}, \bibinfo {author} {\bibfnamefont {D.~M.}\ \bibnamefont
  {Zuckerman}},\ and\ \bibinfo {author} {\bibfnamefont {L.~T.}\ \bibnamefont
  {Chong}},\ }\href@noop {} {\bibfield  {journal} {\bibinfo  {journal} {J.
  Chem. Theory Comput.}\ }\textbf {\bibinfo {volume} {11}},\ \bibinfo {pages}
  {800} (\bibinfo {year} {2015})}\BibitemShut {NoStop}%
\bibitem [{\citenamefont {Bogetti}\ \emph {et~al.}(2019)\citenamefont
  {Bogetti}, \citenamefont {Mosto1an}, \citenamefont {Dickson}, \citenamefont
  {Pratt}, \citenamefont {Saglam}, \citenamefont {Harrison}, \citenamefont
  {Adelman}, \citenamefont {Dudek}, \citenamefont {Torrillo}, \citenamefont
  {DeGrave}, \citenamefont {Adhikari}, \citenamefont {Zwier}, \citenamefont
  {Zuckerman},\ and\ \citenamefont {Chong}}]{Bogetti2019}%
  \BibitemOpen
  \bibfield  {author} {\bibinfo {author} {\bibfnamefont {A.~T.}\ \bibnamefont
  {Bogetti}}, \bibinfo {author} {\bibfnamefont {B.}~\bibnamefont {Mosto1an}},
  \bibinfo {author} {\bibfnamefont {A.}~\bibnamefont {Dickson}}, \bibinfo
  {author} {\bibfnamefont {A.}~\bibnamefont {Pratt}}, \bibinfo {author}
  {\bibfnamefont {A.~S.}\ \bibnamefont {Saglam}}, \bibinfo {author}
  {\bibfnamefont {P.~O.}\ \bibnamefont {Harrison}}, \bibinfo {author}
  {\bibfnamefont {J.~L.}\ \bibnamefont {Adelman}}, \bibinfo {author}
  {\bibfnamefont {M.}~\bibnamefont {Dudek}}, \bibinfo {author} {\bibfnamefont
  {P.~A.}\ \bibnamefont {Torrillo}}, \bibinfo {author} {\bibfnamefont {A.~J.}\
  \bibnamefont {DeGrave}}, \bibinfo {author} {\bibfnamefont {U.}~\bibnamefont
  {Adhikari}}, \bibinfo {author} {\bibfnamefont {M.~C.}\ \bibnamefont {Zwier}},
  \bibinfo {author} {\bibfnamefont {D.~M.}\ \bibnamefont {Zuckerman}},\ and\
  \bibinfo {author} {\bibfnamefont {L.~T.}\ \bibnamefont {Chong}},\ }\href@noop
  {} {\bibfield  {journal} {\bibinfo  {journal} {Living J. Comp. Mol. Sci.}\
  }\textbf {\bibinfo {volume} {1}},\ \bibinfo {pages} {10607} (\bibinfo {year}
  {2019})}\BibitemShut {NoStop}%
\bibitem [{\citenamefont {Russo}\ \emph {et~al.}(2022)\citenamefont {Russo},
  \citenamefont {Zhang}, \citenamefont {Leung}, \citenamefont {Bogetti},
  \citenamefont {Thompson}, \citenamefont {DeGrave}, \citenamefont {Torrillo},
  \citenamefont {Pratt}, \citenamefont {Wong}, \citenamefont {Xia},
  \citenamefont {Copperman}, \citenamefont {Adelman}, \citenamefont {Zwier},
  \citenamefont {LeBard}, \citenamefont {Zuckerman},\ and\ \citenamefont
  {Chong}}]{Russo2022}%
  \BibitemOpen
  \bibfield  {author} {\bibinfo {author} {\bibfnamefont {J.}~\bibnamefont
  {Russo}}, \bibinfo {author} {\bibfnamefont {S.}~\bibnamefont {Zhang}},
  \bibinfo {author} {\bibfnamefont {J.}~\bibnamefont {Leung}}, \bibinfo
  {author} {\bibfnamefont {A.}~\bibnamefont {Bogetti}}, \bibinfo {author}
  {\bibfnamefont {J.}~\bibnamefont {Thompson}}, \bibinfo {author}
  {\bibfnamefont {A.}~\bibnamefont {DeGrave}}, \bibinfo {author} {\bibfnamefont
  {P.}~\bibnamefont {Torrillo}}, \bibinfo {author} {\bibfnamefont
  {A.}~\bibnamefont {Pratt}}, \bibinfo {author} {\bibfnamefont
  {K.}~\bibnamefont {Wong}}, \bibinfo {author} {\bibfnamefont {J.}~\bibnamefont
  {Xia}}, \bibinfo {author} {\bibfnamefont {J.}~\bibnamefont {Copperman}},
  \bibinfo {author} {\bibfnamefont {J.}~\bibnamefont {Adelman}}, \bibinfo
  {author} {\bibfnamefont {M.}~\bibnamefont {Zwier}}, \bibinfo {author}
  {\bibfnamefont {D.}~\bibnamefont {LeBard}}, \bibinfo {author} {\bibfnamefont
  {D.}~\bibnamefont {Zuckerman}},\ and\ \bibinfo {author} {\bibfnamefont
  {L.}~\bibnamefont {Chong}},\ }\href@noop {} {\bibfield  {journal} {\bibinfo
  {journal} {J Chem Theory Comput.}\ }\textbf {\bibinfo {volume} {18}},\
  \bibinfo {pages} {638–649} (\bibinfo {year} {2022})}\BibitemShut {NoStop}%
\bibitem [{\citenamefont {Phillips}\ \emph {et~al.}(2005)\citenamefont
  {Phillips}, \citenamefont {Braun}, \citenamefont {Wang}, \citenamefont
  {Gumbart}, \citenamefont {Tajkhorshid}, \citenamefont {Villa}, \citenamefont
  {Chipot}, \citenamefont {Skeel}, \citenamefont {Kale},\ and\ \citenamefont
  {Schulten}}]{namd}%
  \BibitemOpen
  \bibfield  {author} {\bibinfo {author} {\bibfnamefont {J.}~\bibnamefont
  {Phillips}}, \bibinfo {author} {\bibfnamefont {R.}~\bibnamefont {Braun}},
  \bibinfo {author} {\bibfnamefont {W.}~\bibnamefont {Wang}}, \bibinfo {author}
  {\bibfnamefont {J.}~\bibnamefont {Gumbart}}, \bibinfo {author} {\bibfnamefont
  {E.}~\bibnamefont {Tajkhorshid}}, \bibinfo {author} {\bibfnamefont
  {E.}~\bibnamefont {Villa}}, \bibinfo {author} {\bibfnamefont
  {C.}~\bibnamefont {Chipot}}, \bibinfo {author} {\bibfnamefont {R.~D.}\
  \bibnamefont {Skeel}}, \bibinfo {author} {\bibfnamefont {L.}~\bibnamefont
  {Kale}},\ and\ \bibinfo {author} {\bibfnamefont {K.}~\bibnamefont
  {Schulten}},\ }\href@noop {} {\bibfield  {journal} {\bibinfo  {journal} {J.
  Comput. Chem.}\ }\textbf {\bibinfo {volume} {26}},\ \bibinfo {pages} {1781}
  (\bibinfo {year} {2005})}\BibitemShut {NoStop}%
\bibitem [{\citenamefont {Brooks}\ \emph {et~al.}(2009)\citenamefont {Brooks},
  \citenamefont {III}, \citenamefont {Jr.}, \citenamefont {Nilsson},
  \citenamefont {Petrella}, \citenamefont {Roux}, \citenamefont {Won},
  \citenamefont {Archontis}, \citenamefont {Bartels}, \citenamefont {Boresch},\
  and\ \citenamefont {et~al}}]{charmm}%
  \BibitemOpen
  \bibfield  {author} {\bibinfo {author} {\bibfnamefont {B.~R.}\ \bibnamefont
  {Brooks}}, \bibinfo {author} {\bibfnamefont {C.~L.~B.}\ \bibnamefont {III}},
  \bibinfo {author} {\bibfnamefont {A.~D.~M.}\ \bibnamefont {Jr.}}, \bibinfo
  {author} {\bibfnamefont {L.}~\bibnamefont {Nilsson}}, \bibinfo {author}
  {\bibfnamefont {R.~J.}\ \bibnamefont {Petrella}}, \bibinfo {author}
  {\bibfnamefont {B.}~\bibnamefont {Roux}}, \bibinfo {author} {\bibfnamefont
  {Y.}~\bibnamefont {Won}}, \bibinfo {author} {\bibfnamefont {G.}~\bibnamefont
  {Archontis}}, \bibinfo {author} {\bibfnamefont {C.}~\bibnamefont {Bartels}},
  \bibinfo {author} {\bibfnamefont {S.}~\bibnamefont {Boresch}},\ and\ \bibinfo
  {author} {\bibnamefont {et~al}},\ }\href@noop {} {\bibfield  {journal}
  {\bibinfo  {journal} {J. Comput. Chem.}\ }\textbf {\bibinfo {volume} {30}},\
  \bibinfo {pages} {1545} (\bibinfo {year} {2009})}\BibitemShut {NoStop}%
\bibitem [{\citenamefont {Choe}(2020)}]{Choe2020}%
  \BibitemOpen
  \bibfield  {author} {\bibinfo {author} {\bibfnamefont {S.}~\bibnamefont
  {Choe}},\ }\href@noop {} {\bibfield  {journal} {\bibinfo  {journal} {AIP
  Advances}\ }\textbf {\bibinfo {volume} {10}},\ \bibinfo {pages} {105103}
  (\bibinfo {year} {2020})}\BibitemShut {NoStop}%
\bibitem [{\citenamefont {Jo}\ \emph {et~al.}(2008)\citenamefont {Jo},
  \citenamefont {Kim}, \citenamefont {Iyer},\ and\ \citenamefont
  {Im}}]{charmm-gui}%
  \BibitemOpen
  \bibfield  {author} {\bibinfo {author} {\bibfnamefont {S.}~\bibnamefont
  {Jo}}, \bibinfo {author} {\bibfnamefont {T.}~\bibnamefont {Kim}}, \bibinfo
  {author} {\bibfnamefont {V.~G.}\ \bibnamefont {Iyer}},\ and\ \bibinfo
  {author} {\bibfnamefont {W.}~\bibnamefont {Im}},\ }\href@noop {} {\bibfield
  {journal} {\bibinfo  {journal} {J. Comput. Chem.}\ }\textbf {\bibinfo
  {volume} {29}},\ \bibinfo {pages} {1859} (\bibinfo {year}
  {2008})}\BibitemShut {NoStop}%
\bibitem [{\citenamefont {Walrant}\ \emph {et~al.}(2012)\citenamefont
  {Walrant}, \citenamefont {Vogel}, \citenamefont {Correia}, \citenamefont
  {Lequin}, \citenamefont {Olausson}, \citenamefont {Desbat}, \citenamefont
  {Sagan},\ and\ \citenamefont {Alves}}]{Walrant2012}%
  \BibitemOpen
  \bibfield  {author} {\bibinfo {author} {\bibfnamefont {A.}~\bibnamefont
  {Walrant}}, \bibinfo {author} {\bibfnamefont {A.}~\bibnamefont {Vogel}},
  \bibinfo {author} {\bibfnamefont {I.}~\bibnamefont {Correia}}, \bibinfo
  {author} {\bibfnamefont {O.}~\bibnamefont {Lequin}}, \bibinfo {author}
  {\bibfnamefont {B.~E.~S.}\ \bibnamefont {Olausson}}, \bibinfo {author}
  {\bibfnamefont {B.}~\bibnamefont {Desbat}}, \bibinfo {author} {\bibfnamefont
  {S.}~\bibnamefont {Sagan}},\ and\ \bibinfo {author} {\bibfnamefont {I.~D.}\
  \bibnamefont {Alves}},\ }\href@noop {} {\bibfield  {journal} {\bibinfo
  {journal} {Biochimica Biophysica Acta}\ }\textbf {\bibinfo {volume} {1818}},\
  \bibinfo {pages} {1755} (\bibinfo {year} {2012})}\BibitemShut {NoStop}%
\bibitem [{\citenamefont {Roux}(2008)}]{Roux2008}%
  \BibitemOpen
  \bibfield  {author} {\bibinfo {author} {\bibfnamefont {B.}~\bibnamefont
  {Roux}},\ }\href@noop {} {\bibfield  {journal} {\bibinfo  {journal} {Biophys.
  J.}\ }\textbf {\bibinfo {volume} {95}},\ \bibinfo {pages} {4205} (\bibinfo
  {year} {2008})}\BibitemShut {NoStop}%
\bibitem [{\citenamefont {Michaud-Agrawal}\ \emph {et~al.}(2011)\citenamefont
  {Michaud-Agrawal}, \citenamefont {Denning}, \citenamefont {Woolf},\ and\
  \citenamefont {Beckstein}}]{Michaud2011}%
  \BibitemOpen
  \bibfield  {author} {\bibinfo {author} {\bibfnamefont {N.}~\bibnamefont
  {Michaud-Agrawal}}, \bibinfo {author} {\bibfnamefont {E.~J.}\ \bibnamefont
  {Denning}}, \bibinfo {author} {\bibfnamefont {T.~B.}\ \bibnamefont {Woolf}},\
  and\ \bibinfo {author} {\bibfnamefont {O.}~\bibnamefont {Beckstein}},\
  }\href@noop {} {\bibfield  {journal} {\bibinfo  {journal} {J. Comput. Chem.}\
  }\textbf {\bibinfo {volume} {32}},\ \bibinfo {pages} {2319} (\bibinfo {year}
  {2011})}\BibitemShut {NoStop}%
\bibitem [{\citenamefont {Grossfield}(2022)}]{Grossfield2022}%
  \BibitemOpen
  \bibfield  {author} {\bibinfo {author} {\bibfnamefont {A.}~\bibnamefont
  {Grossfield}},\ }\href@noop {} {\bibfield  {journal} {\bibinfo  {journal}
  {~~~}\ } (\bibinfo {year} {2022})}\BibitemShut {NoStop}%
\bibitem [{\citenamefont {Rothbard}\ \emph {et~al.}(2004)\citenamefont
  {Rothbard}, \citenamefont {Jessop}, \citenamefont {Lewis}, \citenamefont
  {Murray},\ and\ \citenamefont {Wender}}]{Rothbard2004}%
  \BibitemOpen
  \bibfield  {author} {\bibinfo {author} {\bibfnamefont {J.}~\bibnamefont
  {Rothbard}}, \bibinfo {author} {\bibfnamefont {T.}~\bibnamefont {Jessop}},
  \bibinfo {author} {\bibfnamefont {R.}~\bibnamefont {Lewis}}, \bibinfo
  {author} {\bibfnamefont {B.}~\bibnamefont {Murray}},\ and\ \bibinfo {author}
  {\bibfnamefont {P.}~\bibnamefont {Wender}},\ }\href@noop {} {\bibfield
  {journal} {\bibinfo  {journal} {J Am Chem Soc}\ }\textbf {\bibinfo {volume}
  {126}},\ \bibinfo {pages} {9506–9507} (\bibinfo {year} {2004})}\BibitemShut
  {NoStop}%
\bibitem [{\citenamefont {Tang}\ \emph {et~al.}(2008)\citenamefont {Tang},
  \citenamefont {Waring}, \citenamefont {Lehrer},\ and\ \citenamefont
  {Hong}}]{Tang2008}%
  \BibitemOpen
  \bibfield  {author} {\bibinfo {author} {\bibfnamefont {M.}~\bibnamefont
  {Tang}}, \bibinfo {author} {\bibfnamefont {A.}~\bibnamefont {Waring}},
  \bibinfo {author} {\bibfnamefont {R.}~\bibnamefont {Lehrer}},\ and\ \bibinfo
  {author} {\bibfnamefont {M.}~\bibnamefont {Hong}},\ }\href@noop {} {\bibfield
   {journal} {\bibinfo  {journal} {Angewandte Chemie Int Ed.}\ }\textbf
  {\bibinfo {volume} {47}},\ \bibinfo {pages} {3202–3205} (\bibinfo {year}
  {2008})}\BibitemShut {NoStop}%
\bibitem [{\citenamefont {Takechi-Haraya}\ and\ \citenamefont
  {Saito}(2018)}]{Takechi-Haraya2018}%
  \BibitemOpen
  \bibfield  {author} {\bibinfo {author} {\bibfnamefont {Y.}~\bibnamefont
  {Takechi-Haraya}}\ and\ \bibinfo {author} {\bibfnamefont {H.}~\bibnamefont
  {Saito}},\ }\href@noop {} {\bibfield  {journal} {\bibinfo  {journal} {Current
  Protein and Peptide Science}\ }\textbf {\bibinfo {volume} {19}},\ \bibinfo
  {pages} {623} (\bibinfo {year} {2018})}\BibitemShut {NoStop}%
\bibitem [{\citenamefont {Pei}(2022)}]{Pei2022}%
  \BibitemOpen
  \bibfield  {author} {\bibinfo {author} {\bibfnamefont {D.}~\bibnamefont
  {Pei}},\ }\href@noop {} {\bibfield  {journal} {\bibinfo  {journal} {Accounts
  Chem Res}\ }\textbf {\bibinfo {volume} {55}},\ \bibinfo {pages} {309}
  (\bibinfo {year} {2022})}\BibitemShut {NoStop}%
\bibitem [{\citenamefont {Eir\'{i}ksd\'{o}ttir}\ \emph
  {et~al.}(2010)\citenamefont {Eir\'{i}ksd\'{o}ttir}, \citenamefont {Konate},
  \citenamefont {Langel}, \citenamefont {Divita},\ and\ \citenamefont
  {Deshayes}}]{Eiriksdottir2010}%
  \BibitemOpen
  \bibfield  {author} {\bibinfo {author} {\bibfnamefont {E.}~\bibnamefont
  {Eir\'{i}ksd\'{o}ttir}}, \bibinfo {author} {\bibfnamefont {K.}~\bibnamefont
  {Konate}}, \bibinfo {author} {\bibfnamefont {{\"U}.}~\bibnamefont {Langel}},
  \bibinfo {author} {\bibfnamefont {G.}~\bibnamefont {Divita}},\ and\ \bibinfo
  {author} {\bibfnamefont {S.}~\bibnamefont {Deshayes}},\ }\href@noop {}
  {\bibfield  {journal} {\bibinfo  {journal} {Biochimica Biophysica Acta}\
  }\textbf {\bibinfo {volume} {1798}},\ \bibinfo {pages} {1119} (\bibinfo
  {year} {2010})}\BibitemShut {NoStop}%
\bibitem [{\citenamefont {Humphrey}\ \emph {et~al.}(1996)\citenamefont
  {Humphrey}, \citenamefont {Dalke},\ and\ \citenamefont {Schulten}}]{vmd}%
  \BibitemOpen
  \bibfield  {author} {\bibinfo {author} {\bibfnamefont {W.}~\bibnamefont
  {Humphrey}}, \bibinfo {author} {\bibfnamefont {A.}~\bibnamefont {Dalke}},\
  and\ \bibinfo {author} {\bibfnamefont {K.}~\bibnamefont {Schulten}},\
  }\href@noop {} {\bibfield  {journal} {\bibinfo  {journal} {J. Molec.
  Graphics}\ }\textbf {\bibinfo {volume} {14}},\ \bibinfo {pages} {33}
  (\bibinfo {year} {1996})}\BibitemShut {NoStop}%
\bibitem [{\citenamefont {Sengupta}\ \emph {et~al.}(2008)\citenamefont
  {Sengupta}, \citenamefont {Leontiadou}, \citenamefont {Mark},\ and\
  \citenamefont {Marrink}}]{Sengupta2008}%
  \BibitemOpen
  \bibfield  {author} {\bibinfo {author} {\bibfnamefont {D.}~\bibnamefont
  {Sengupta}}, \bibinfo {author} {\bibfnamefont {H.}~\bibnamefont
  {Leontiadou}}, \bibinfo {author} {\bibfnamefont {A.~E.}\ \bibnamefont
  {Mark}},\ and\ \bibinfo {author} {\bibfnamefont {S.-J.}\ \bibnamefont
  {Marrink}},\ }\href@noop {} {\bibfield  {journal} {\bibinfo  {journal}
  {Biochimica et Biophysica Acta}\ }\textbf {\bibinfo {volume} {1778}},\
  \bibinfo {pages} {2308} (\bibinfo {year} {2008})}\BibitemShut {NoStop}%
\bibitem [{\citenamefont {Bennett}\ \emph {et~al.}(2014)\citenamefont
  {Bennett}, \citenamefont {Sapay},\ and\ \citenamefont
  {Tieleman}}]{Bennett2014}%
  \BibitemOpen
  \bibfield  {author} {\bibinfo {author} {\bibfnamefont {W.~F.~D.}\
  \bibnamefont {Bennett}}, \bibinfo {author} {\bibfnamefont {N.}~\bibnamefont
  {Sapay}},\ and\ \bibinfo {author} {\bibfnamefont {D.~P.}\ \bibnamefont
  {Tieleman}},\ }\href@noop {} {\bibfield  {journal} {\bibinfo  {journal}
  {Biophys. J.}\ }\textbf {\bibinfo {volume} {106}},\ \bibinfo {pages} {210}
  (\bibinfo {year} {2014})}\BibitemShut {NoStop}%
\bibitem [{\citenamefont {Huang}\ and\ \citenamefont
  {Garcia}(2013)}]{Huang2013}%
  \BibitemOpen
  \bibfield  {author} {\bibinfo {author} {\bibfnamefont {K.}~\bibnamefont
  {Huang}}\ and\ \bibinfo {author} {\bibfnamefont {A.~E.}\ \bibnamefont
  {Garcia}},\ }\href@noop {} {\bibfield  {journal} {\bibinfo  {journal}
  {Biophys. J.}\ }\textbf {\bibinfo {volume} {104}},\ \bibinfo {pages} {412}
  (\bibinfo {year} {2013})}\BibitemShut {NoStop}%
\bibitem [{\citenamefont {Marrink}\ \emph {et~al.}(2001)\citenamefont
  {Marrink}, \citenamefont {Lindahl}, \citenamefont {Edholm},\ and\
  \citenamefont {Mark}}]{Marrink2001}%
  \BibitemOpen
  \bibfield  {author} {\bibinfo {author} {\bibfnamefont {S.}~\bibnamefont
  {Marrink}}, \bibinfo {author} {\bibfnamefont {E.}~\bibnamefont {Lindahl}},
  \bibinfo {author} {\bibfnamefont {O.}~\bibnamefont {Edholm}},\ and\ \bibinfo
  {author} {\bibfnamefont {A.}~\bibnamefont {Mark}},\ }\href@noop {} {\bibfield
   {journal} {\bibinfo  {journal} {J Am Chem Soc.}\ }\textbf {\bibinfo {volume}
  {123}},\ \bibinfo {pages} {8638–8639} (\bibinfo {year} {2001})}\BibitemShut
  {NoStop}%
\bibitem [{\citenamefont {de~Vries}\ \emph {et~al.}(2004)\citenamefont
  {de~Vries}, \citenamefont {Mark},\ and\ \citenamefont
  {Marrink}}]{deVries2004}%
  \BibitemOpen
  \bibfield  {author} {\bibinfo {author} {\bibfnamefont {A.}~\bibnamefont
  {de~Vries}}, \bibinfo {author} {\bibfnamefont {A.}~\bibnamefont {Mark}},\
  and\ \bibinfo {author} {\bibfnamefont {S.}~\bibnamefont {Marrink}},\
  }\href@noop {} {\bibfield  {journal} {\bibinfo  {journal} {J Am Chem Soc.}\
  }\textbf {\bibinfo {volume} {126}},\ \bibinfo {pages} {4488–4489} (\bibinfo
  {year} {2004})}\BibitemShut {NoStop}%
\bibitem [{\citenamefont {Fattal}\ \emph {et~al.}(1994)\citenamefont {Fattal},
  \citenamefont {Nir}, \citenamefont {Parente},\ and\ \citenamefont
  {F.~C.~Szoka}}]{Fattal1994}%
  \BibitemOpen
  \bibfield  {author} {\bibinfo {author} {\bibfnamefont {E.}~\bibnamefont
  {Fattal}}, \bibinfo {author} {\bibfnamefont {S.}~\bibnamefont {Nir}},
  \bibinfo {author} {\bibfnamefont {R.~A.}\ \bibnamefont {Parente}},\ and\
  \bibinfo {author} {\bibfnamefont {J.}~\bibnamefont {F.~C.~Szoka}},\
  }\href@noop {} {\bibfield  {journal} {\bibinfo  {journal} {Biochemistry}\
  }\textbf {\bibinfo {volume} {33}},\ \bibinfo {pages} {6721} (\bibinfo {year}
  {1994})}\BibitemShut {NoStop}%
\bibitem [{\citenamefont {Matsuzaki}\ \emph {et~al.}(1996)\citenamefont
  {Matsuzaki}, \citenamefont {Murase}, \citenamefont {Fujii},\ and\
  \citenamefont {Miyajima}}]{Matsuzaki1996}%
  \BibitemOpen
  \bibfield  {author} {\bibinfo {author} {\bibfnamefont {K.}~\bibnamefont
  {Matsuzaki}}, \bibinfo {author} {\bibfnamefont {O.}~\bibnamefont {Murase}},
  \bibinfo {author} {\bibfnamefont {N.}~\bibnamefont {Fujii}},\ and\ \bibinfo
  {author} {\bibfnamefont {K.}~\bibnamefont {Miyajima}},\ }\href@noop {}
  {\bibfield  {journal} {\bibinfo  {journal} {Biochemistry}\ }\textbf {\bibinfo
  {volume} {35}},\ \bibinfo {pages} {11361} (\bibinfo {year}
  {1996})}\BibitemShut {NoStop}%
\bibitem [{\citenamefont {Kol}\ \emph {et~al.}(2003)\citenamefont {Kol},
  \citenamefont {van Dalen}, \citenamefont {de~Kroon},\ and\ \citenamefont
  {de~Kruijff}}]{Kol2003}%
  \BibitemOpen
  \bibfield  {author} {\bibinfo {author} {\bibfnamefont {M.~A.}\ \bibnamefont
  {Kol}}, \bibinfo {author} {\bibfnamefont {A.}~\bibnamefont {van Dalen}},
  \bibinfo {author} {\bibfnamefont {A.~I. P.~M.}\ \bibnamefont {de~Kroon}},\
  and\ \bibinfo {author} {\bibfnamefont {B.}~\bibnamefont {de~Kruijff}},\
  }\href@noop {} {\bibfield  {journal} {\bibinfo  {journal} {The Journal of
  Biological Chemistry}\ }\textbf {\bibinfo {volume} {278}},\ \bibinfo {pages}
  {24586} (\bibinfo {year} {2003})}\BibitemShut {NoStop}%
\bibitem [{\citenamefont {Tieleman}\ and\ \citenamefont
  {Marrink}(2006)}]{Tieleman2006}%
  \BibitemOpen
  \bibfield  {author} {\bibinfo {author} {\bibfnamefont {D.~P.}\ \bibnamefont
  {Tieleman}}\ and\ \bibinfo {author} {\bibfnamefont {S.-J.}\ \bibnamefont
  {Marrink}},\ }\href@noop {} {\bibfield  {journal} {\bibinfo  {journal} {J.
  Am. Chem. Soc.}\ }\textbf {\bibinfo {volume} {128}},\ \bibinfo {pages}
  {12462} (\bibinfo {year} {2006})}\BibitemShut {NoStop}%
\bibitem [{\citenamefont {Gurtovenko}\ and\ \citenamefont
  {Vattulainen}(2007)}]{Gurtovenko2007}%
  \BibitemOpen
  \bibfield  {author} {\bibinfo {author} {\bibfnamefont {A.~A.}\ \bibnamefont
  {Gurtovenko}}\ and\ \bibinfo {author} {\bibfnamefont {I.}~\bibnamefont
  {Vattulainen}},\ }\href@noop {} {\bibfield  {journal} {\bibinfo  {journal}
  {J. Phys. Chem. B}\ }\textbf {\bibinfo {volume} {111}},\ \bibinfo {pages}
  {13554} (\bibinfo {year} {2007})}\BibitemShut {NoStop}%
\bibitem [{\citenamefont {Parisio}\ \emph {et~al.}(2016)\citenamefont
  {Parisio}, \citenamefont {Ferrarini},\ and\ \citenamefont
  {Sperotto}}]{Parosio2016}%
  \BibitemOpen
  \bibfield  {author} {\bibinfo {author} {\bibfnamefont {G.}~\bibnamefont
  {Parisio}}, \bibinfo {author} {\bibfnamefont {A.}~\bibnamefont {Ferrarini}},\
  and\ \bibinfo {author} {\bibfnamefont {M.~M.}\ \bibnamefont {Sperotto}},\
  }\href@noop {} {\bibfield  {journal} {\bibinfo  {journal} {International
  Journal of Advances in Engineering Sciences and Applied Mathematics}\
  }\textbf {\bibinfo {volume} {8}},\ \bibinfo {pages} {134} (\bibinfo {year}
  {2016})}\BibitemShut {NoStop}%
\bibitem [{\citenamefont {Allhusen}\ and\ \citenamefont
  {Conboy}(2017)}]{Allhusen2017}%
  \BibitemOpen
  \bibfield  {author} {\bibinfo {author} {\bibfnamefont {J.~S.}\ \bibnamefont
  {Allhusen}}\ and\ \bibinfo {author} {\bibfnamefont {J.~C.}\ \bibnamefont
  {Conboy}},\ }\href@noop {} {\bibfield  {journal} {\bibinfo  {journal} {Acc.
  Chem. Res.}\ }\textbf {\bibinfo {volume} {50}},\ \bibinfo {pages} {58}
  (\bibinfo {year} {2017})}\BibitemShut {NoStop}%
\bibitem [{\citenamefont {Nguyen}\ \emph {et~al.}(2021)\citenamefont {Nguyen},
  \citenamefont {DiPasquale}, \citenamefont {Rickeard}, \citenamefont {Yip},
  \citenamefont {Greco}, \citenamefont {Kelley},\ and\ \citenamefont
  {Marquardt}}]{Nguyen2021}%
  \BibitemOpen
  \bibfield  {author} {\bibinfo {author} {\bibfnamefont {M.~H.~L.}\
  \bibnamefont {Nguyen}}, \bibinfo {author} {\bibfnamefont {M.}~\bibnamefont
  {DiPasquale}}, \bibinfo {author} {\bibfnamefont {B.~W.}\ \bibnamefont
  {Rickeard}}, \bibinfo {author} {\bibfnamefont {C.~G.}\ \bibnamefont {Yip}},
  \bibinfo {author} {\bibfnamefont {K.~N.}\ \bibnamefont {Greco}}, \bibinfo
  {author} {\bibfnamefont {E.~G.}\ \bibnamefont {Kelley}},\ and\ \bibinfo
  {author} {\bibfnamefont {D.}~\bibnamefont {Marquardt}},\ }\href@noop {}
  {\bibfield  {journal} {\bibinfo  {journal} {New J. Chem.}\ }\textbf {\bibinfo
  {volume} {45}},\ \bibinfo {pages} {447} (\bibinfo {year} {2021})}\BibitemShut
  {NoStop}%
\bibitem [{\citenamefont {Torrillo}\ \emph {et~al.}(2021)\citenamefont
  {Torrillo}, \citenamefont {Bogetti},\ and\ \citenamefont
  {Chong}}]{Torrillo2021}%
  \BibitemOpen
  \bibfield  {author} {\bibinfo {author} {\bibfnamefont {P.}~\bibnamefont
  {Torrillo}}, \bibinfo {author} {\bibfnamefont {A.}~\bibnamefont {Bogetti}},\
  and\ \bibinfo {author} {\bibfnamefont {L.}~\bibnamefont {Chong}},\
  }\href@noop {} {\bibfield  {journal} {\bibinfo  {journal} {J Phys Chem.}\
  }\textbf {\bibinfo {volume} {125}},\ \bibinfo {pages} {1642–1649} (\bibinfo
  {year} {2021})}\BibitemShut {NoStop}%
\bibitem [{\citenamefont {Huang}(2000)}]{Huang2000}%
  \BibitemOpen
  \bibfield  {author} {\bibinfo {author} {\bibfnamefont {H.}~\bibnamefont
  {Huang}},\ }\href@noop {} {\bibfield  {journal} {\bibinfo  {journal}
  {Biochemistry}\ }\textbf {\bibinfo {volume} {39}},\ \bibinfo {pages} {8347}
  (\bibinfo {year} {2000})}\BibitemShut {NoStop}%
\bibitem [{\citenamefont {Huang}\ \emph {et~al.}(2004)\citenamefont {Huang},
  \citenamefont {Chen},\ and\ \citenamefont {Lee}}]{Huang2004}%
  \BibitemOpen
  \bibfield  {author} {\bibinfo {author} {\bibfnamefont {H.}~\bibnamefont
  {Huang}}, \bibinfo {author} {\bibfnamefont {F.-Y.}\ \bibnamefont {Chen}},\
  and\ \bibinfo {author} {\bibfnamefont {M.-T.}\ \bibnamefont {Lee}},\
  }\href@noop {} {\bibfield  {journal} {\bibinfo  {journal} {Phys Rev Lett.}\
  }\textbf {\bibinfo {volume} {92}},\ \bibinfo {pages} {198304} (\bibinfo
  {year} {2004})}\BibitemShut {NoStop}%
\bibitem [{\citenamefont {Wang}\ \emph {et~al.}(2018)\citenamefont {Wang},
  \citenamefont {Zhang}, \citenamefont {Zhang}, \citenamefont {Mao},
  \citenamefont {Lu},\ and\ \citenamefont {Liu}}]{Wang2018}%
  \BibitemOpen
  \bibfield  {author} {\bibinfo {author} {\bibfnamefont {B.}~\bibnamefont
  {Wang}}, \bibinfo {author} {\bibfnamefont {J.}~\bibnamefont {Zhang}},
  \bibinfo {author} {\bibfnamefont {Y.}~\bibnamefont {Zhang}}, \bibinfo
  {author} {\bibfnamefont {Z.}~\bibnamefont {Mao}}, \bibinfo {author}
  {\bibfnamefont {N.}~\bibnamefont {Lu}},\ and\ \bibinfo {author}
  {\bibfnamefont {Q.~H.}\ \bibnamefont {Liu}},\ }\href@noop {} {\bibfield
  {journal} {\bibinfo  {journal} {RSC Adv.}\ }\textbf {\bibinfo {volume} {8}},\
  \bibinfo {pages} {41517} (\bibinfo {year} {2018})}\BibitemShut {NoStop}%
\bibitem [{\citenamefont {Via}\ \emph {et~al.}(2018)\citenamefont {Via},
  \citenamefont {Klug}, \citenamefont {Wilke}, \citenamefont {Mayorga},\ and\
  \citenamefont {Popolo}}]{Via2018}%
  \BibitemOpen
  \bibfield  {author} {\bibinfo {author} {\bibfnamefont {M.~A.}\ \bibnamefont
  {Via}}, \bibinfo {author} {\bibfnamefont {J.}~\bibnamefont {Klug}}, \bibinfo
  {author} {\bibfnamefont {N.}~\bibnamefont {Wilke}}, \bibinfo {author}
  {\bibfnamefont {L.~S.}\ \bibnamefont {Mayorga}},\ and\ \bibinfo {author}
  {\bibfnamefont {M.~G.~D.}\ \bibnamefont {Popolo}},\ }\href@noop {} {\bibfield
   {journal} {\bibinfo  {journal} {Phys. Chem. Chem. Phys.}\ }\textbf {\bibinfo
  {volume} {20}},\ \bibinfo {pages} {5180} (\bibinfo {year}
  {2018})}\BibitemShut {NoStop}%
\end{thebibliography}%


\begin{thebibliography}{3}
\expandafter\ifx\csname natexlab\endcsname\relax\def\natexlab#1{#1}\fi
\expandafter\ifx\csname bibnamefont\endcsname\relax
  \def\bibnamefont#1{#1}\fi
\expandafter\ifx\csname bibfnamefont\endcsname\relax
  \def\bibfnamefont#1{#1}\fi
\expandafter\ifx\csname citenamefont\endcsname\relax
  \def\citenamefont#1{#1}\fi
\expandafter\ifx\csname url\endcsname\relax
  \def\url#1{\texttt{#1}}\fi
\expandafter\ifx\csname urlprefix\endcsname\relax\def\urlprefix{URL }\fi
\providecommand{\bibinfo}[2]{#2}
\providecommand{\eprint}[2][]{\url{#2}}

\bibitem[{\citenamefont{Humphrey et~al.}(1996)\citenamefont{Humphrey, Dalke,
  and Schulten}}]{vmd}
\bibinfo{author}{\bibfnamefont{W.}~\bibnamefont{Humphrey}},
  \bibinfo{author}{\bibfnamefont{A.}~\bibnamefont{Dalke}}, \bibnamefont{and}
  \bibinfo{author}{\bibfnamefont{K.}~\bibnamefont{Schulten}},
  \bibinfo{journal}{J. Molec. Graphics} \textbf{\bibinfo{volume}{14}},
  \bibinfo{pages}{33} (\bibinfo{year}{1996}).

\bibitem[{\citenamefont{Choe}(2020)}]{Choe2020}
\bibinfo{author}{\bibfnamefont{S.}~\bibnamefont{Choe}}, \bibinfo{journal}{AIP
  Advances} \textbf{\bibinfo{volume}{10}}, \bibinfo{pages}{105103}
  (\bibinfo{year}{2020}).

\bibitem[{\citenamefont{Choe}(2021)}]{Choe2021}
\bibinfo{author}{\bibfnamefont{S.}~\bibnamefont{Choe}},
  \bibinfo{journal}{Membranes} \textbf{\bibinfo{volume}{11}},
  \bibinfo{pages}{974} (\bibinfo{year}{2021}).

\end{thebibliography}

\end{document}


\begin{center}
{\LARGE {Supplementary Information}}
\end{center}

\vspace{1cm}
\title{Translocation of a single Arg$_9$ peptide across a DOPC/DOPG(4:1) model membrane using the weighted ensemble method}
\author{Seungho Choe$^1,^2$}
\address{
$^{1}$ Department of Energy Science \& Engineering, DGIST, Daegu 42988, South Korea\\
$^{2}$ Energy Science \& Engineering Research Center, DGIST, Daegu 42988, South Korea}

%

\maketitle

\newpage

\section{The number of hydrogen bonds between Arg$_9$ and the lipids molecules and the conformational change of Arg$_9$ during the translocation}

\begin{figure}[ht!]
\centering
\includegraphics[width=10cm,height=10cm]{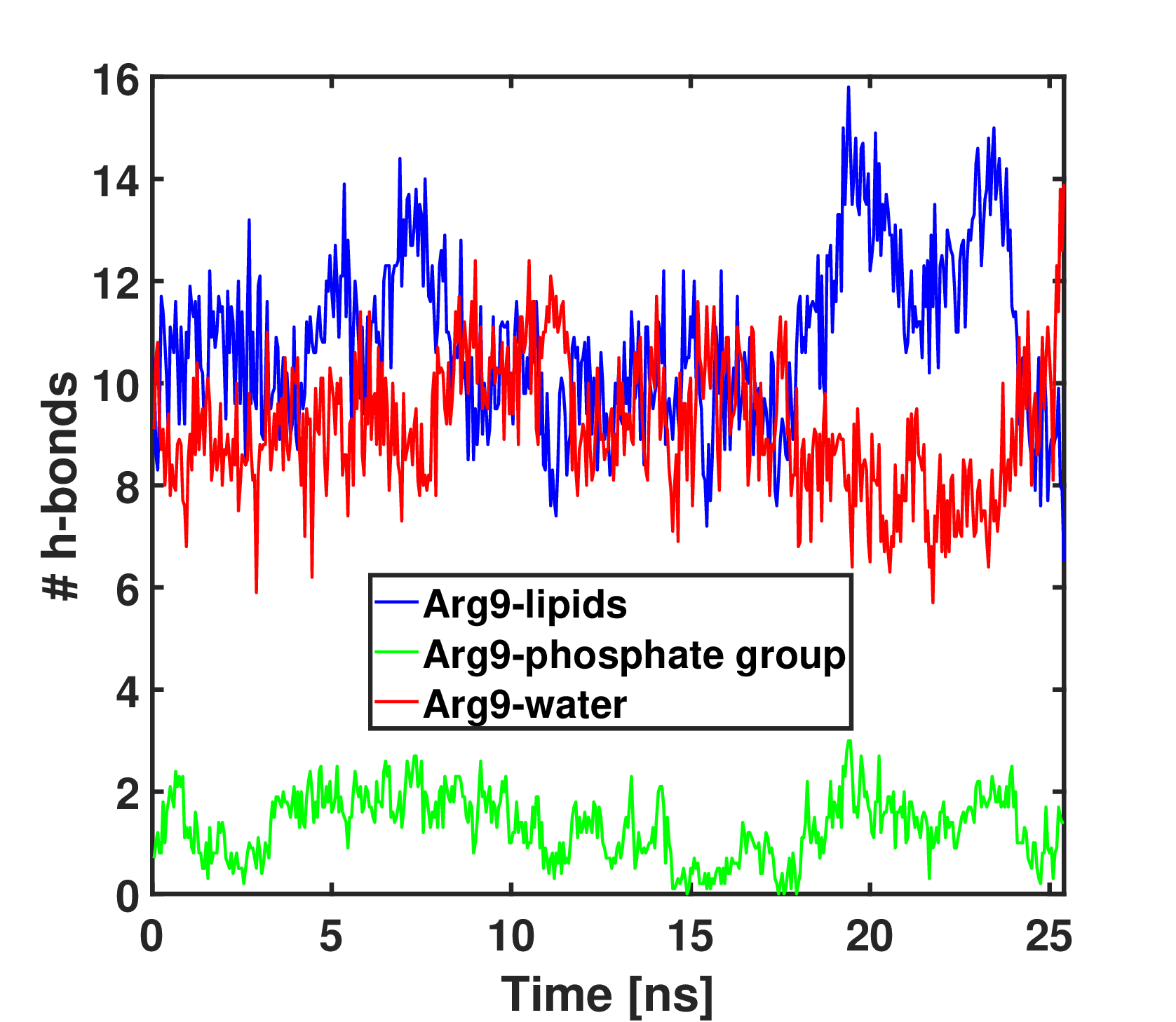}

\caption{The number of hydrogen bonds (blue: between Arg$_9$ and the lipids molecules (including the phosphate group), green: between Arg$_9$ and only the phosphate group, red: between Arg$_9$ and water) vs. the simulation time}\label{figS1}
\end{figure}

We counted the number of hydrogen bonds between Arg$_9$ and the lipid molecules during the WE simulations. Fig. \ref{figS1} denotes the number of hydrogen bonds calculated in VMD \cite{vmd}. The blue line is the number of the hydrogen bonds between Arg$_9$ and the lipids (including the phosphate group), while the green line is between Arg$_9$ and only the phosphate group. On the other hand, the red line denotes the hydrogen bonds between Arg$_9$ and water. Each point in the figure corresponds to an averaged value over 50 ps. The red line shows a rapid increase when Arg$_9$ reaches the bottom of the membrane. The figure shows that the average number of hydrogen bonds between Arg$_9$ and both the lipids and water was about 20. We suggest in the main text that the water flow and the orientation angle of Arg$_9$ can play a role in  the translocation of Arg$_9$. 

\begin{figure}[ht!]
\centering
\includegraphics[width=12cm,height=8cm]{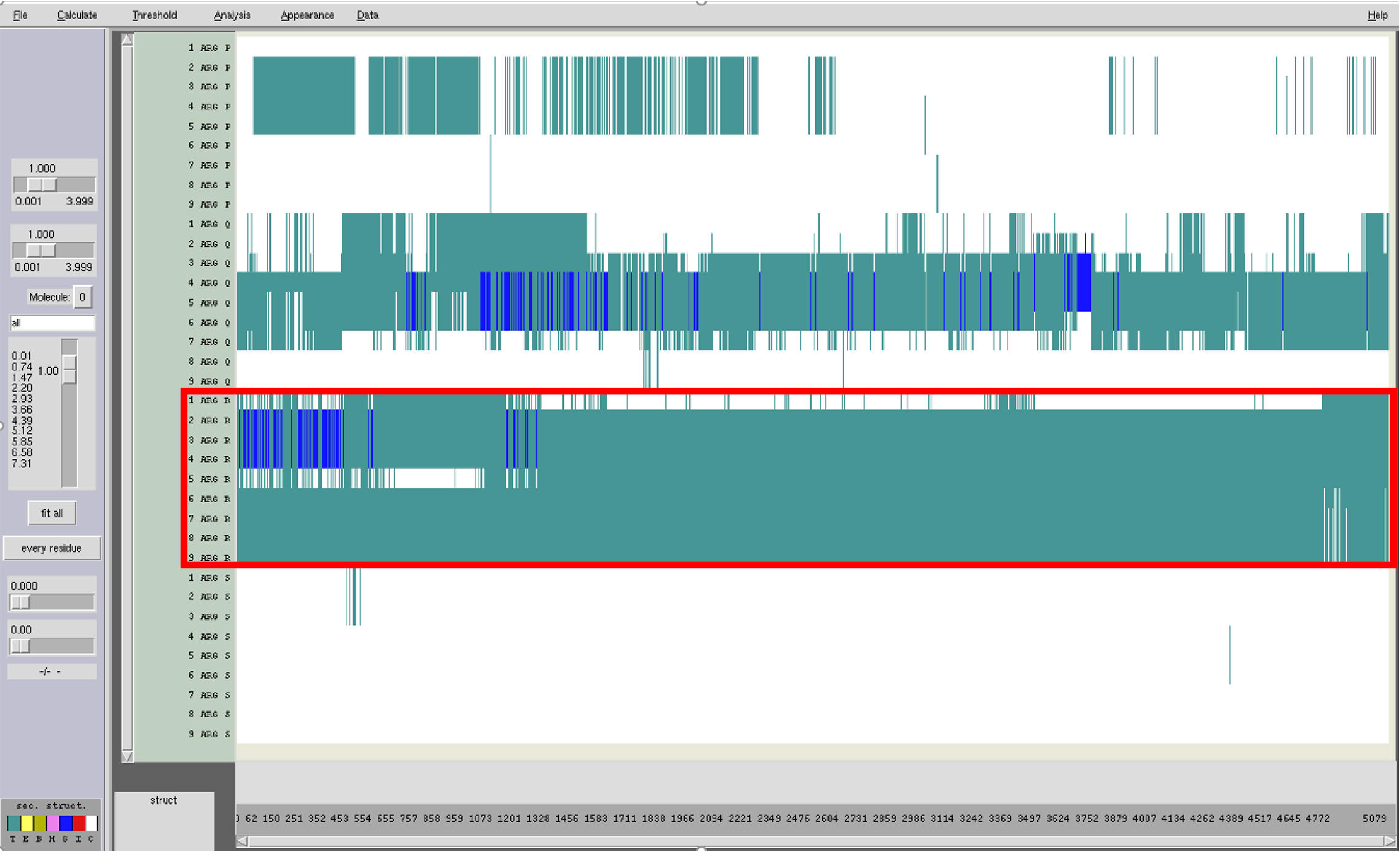}
\caption{The conformational change of Arg$_9$s during the WE simulation. We used the Timeline in VMD \cite{vmd}. The third Arg$_9$ is the one that showed the translocation across the membrane}\label{figS2}
\end{figure}

Fig. \ref{figS2} shows the conformational change of Arg$_9$s during the WE simulation calculated in the Timeline in VMD \cite{vmd}. The third Arg$_9$ is the one that showed the translocation across the membrane, and most of the residues maintained ``turn" conformation during the whole WE simulation.    
The other three Arg$_9$s showed the conformation of the random coil. 

\section{Comparison of the orientation angle of Arg$_9$ at the last part of the translocation}

\begin{figure}[ht!]
\centering
\includegraphics[width=10cm,height=10cm]{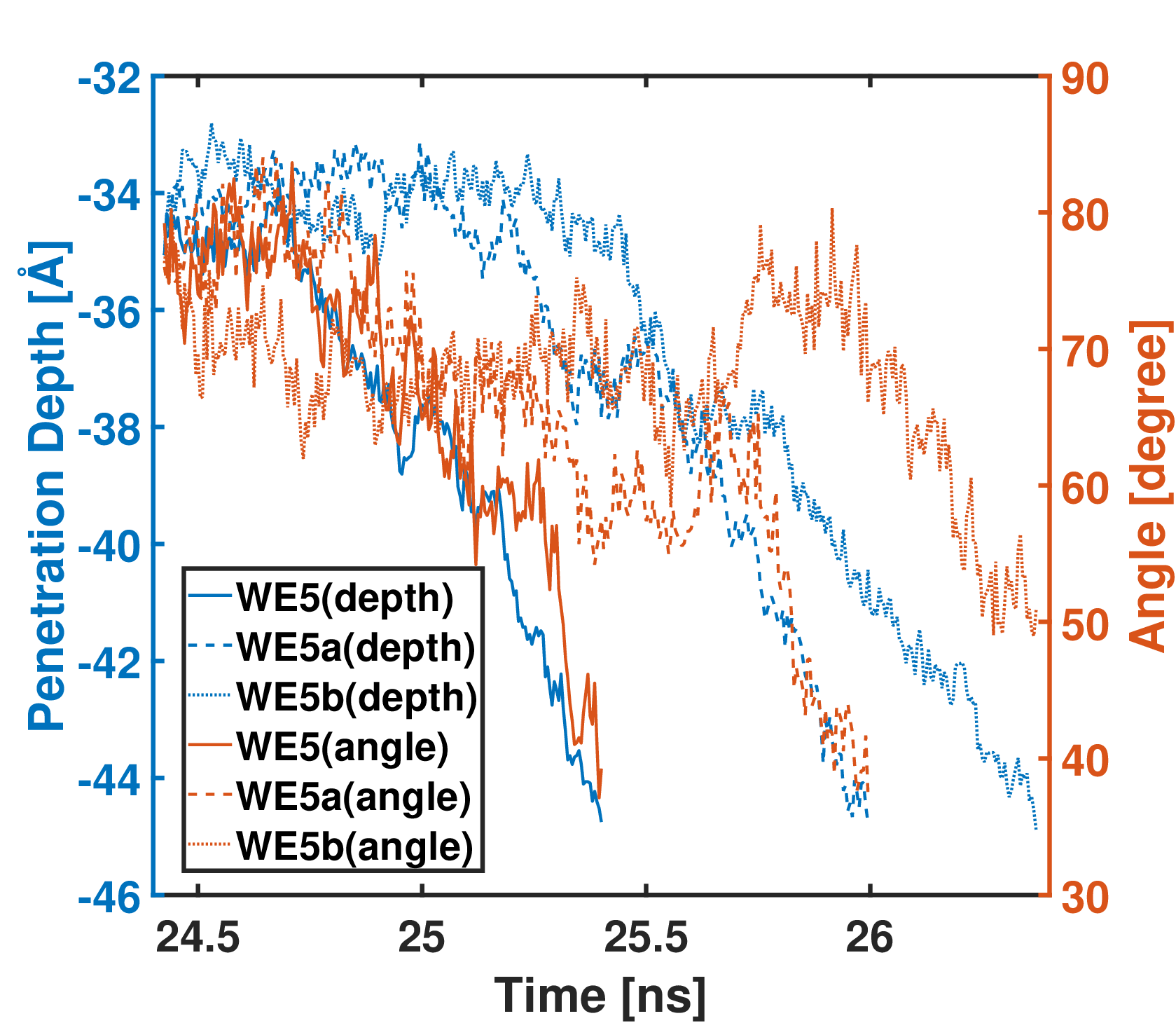}

\caption{The penetration depth and the orientation angle of Arg$_9$ from three different simulations: WE5, WE5a, and WE5b (the last part of the translocation) vs. the simulation time. The time was counted since the WE1 simulation started (see Table 1 in the main text).}\label{figS3}
\end{figure}

In addition to the WE5 simulation, we performed two more simulations (WE5a and WE5b) to compare the orientation angle of Arg$_9$ during the last part of the translocation to the bottom of the membrane. We used the same system set up in the WE5 simulation for the WE5a and WE5b simulations. Therefore, they have the same boundary and bin size. Table S1 denotes a list of three WE simulations performed at the end of translocation, including the WE5 simulation in the main text. 
Fig. \ref{figS3} shows the penetration depth vs. the orientation angle from the WE5, WE5a, and WE5b simulations. The figure shows a strong correlation between the penetration depth and the orientation angle. Arg$_9$ can penetrate rapidly with a smaller orientation angle.

\begin{table}[ht!]
\begin{center}
\caption{A list of three WE simulations performed at the last part of the translocation}\label{tab1}%
\begin{tabular}{@{}cccc@{}}
\hline
simulation no.  & bin boundaries  &  bin size (the smallest) & \ \# iterations \\
\hline
WE5    & [$-$45.0 \AA, $-$33.0 \AA] & 0.10 \AA   & 196 \\
WE5a    & [$-$45.0 \AA, $-$33.0 \AA] & 0.10 \AA   & 315 \\
WE5b    & [$-$45.0 \AA, $-$33.0 \AA] & 0.10 \AA   & 390 \\
\hline
\end{tabular}
\end{center}
\end{table}

%

\section{An additional conventional MD simulation without the WE method}

\begin{figure}[ht!]
\centering
\subfloat[20 ns]{\includegraphics[width=5cm,height=5cm]{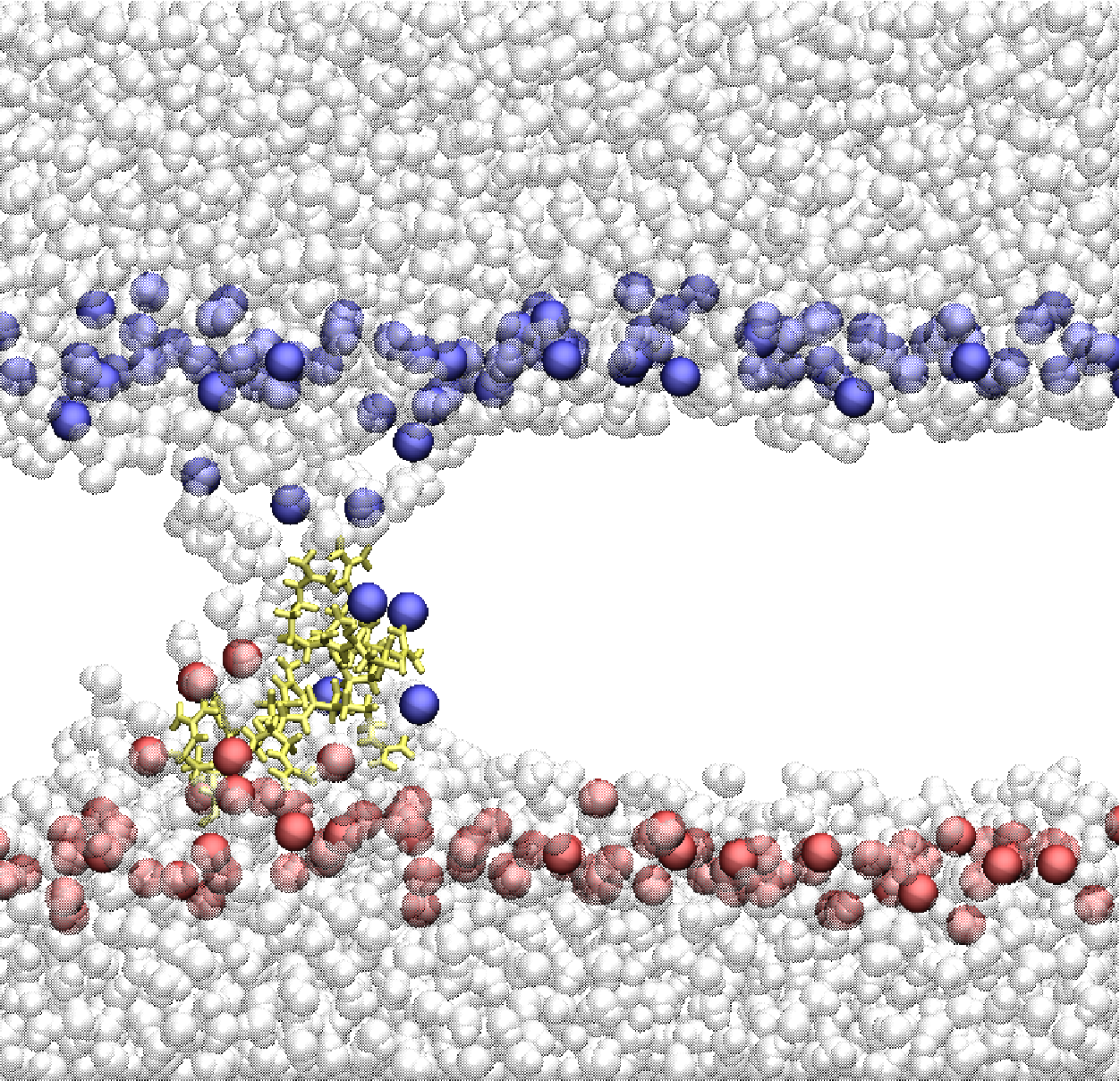}}\vspace{5mm}
\subfloat[40 ns]{\includegraphics[width=5cm,height=5cm]{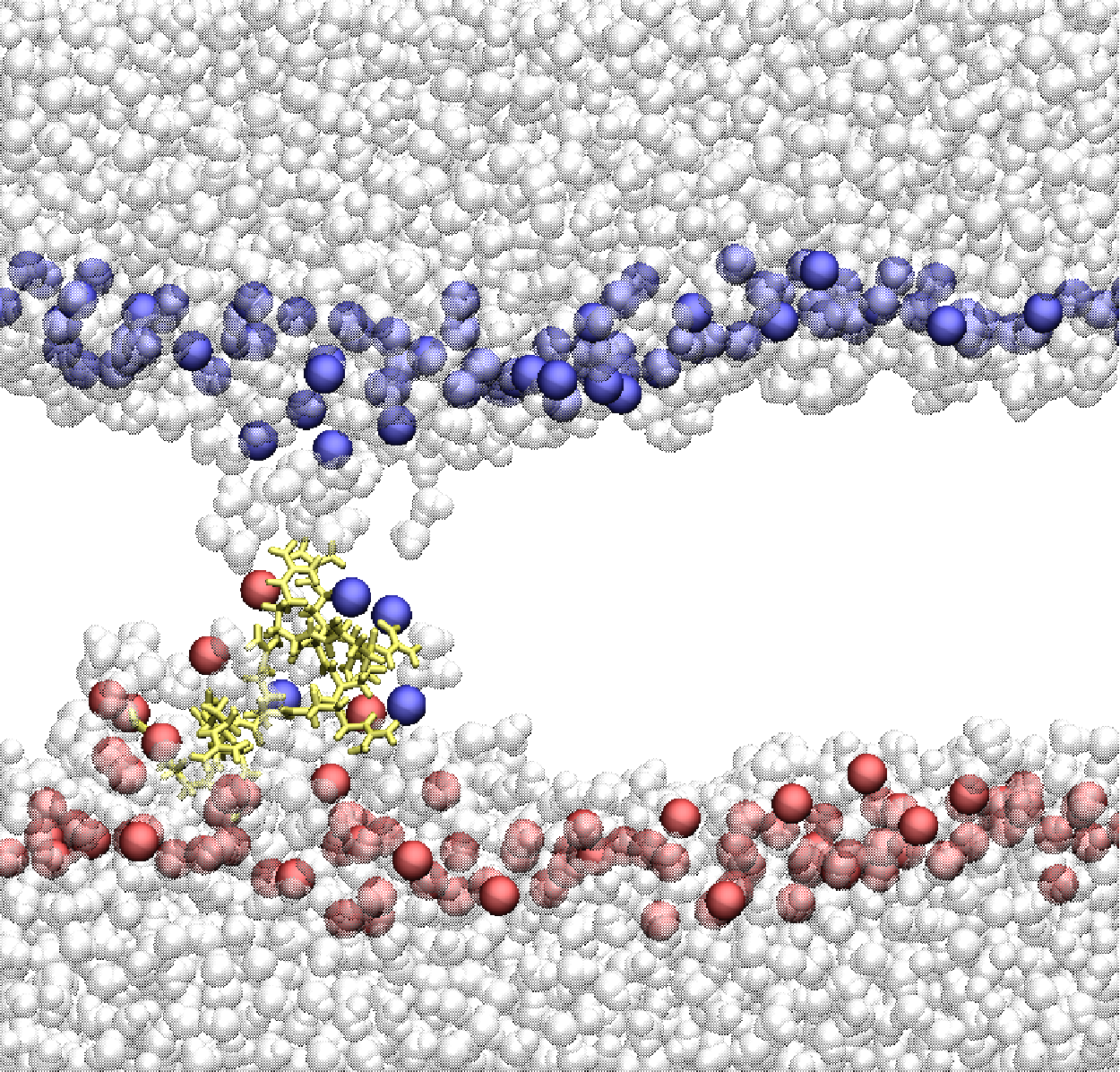}}\vspace{5mm}
\subfloat[60 ns]{\includegraphics[width=5cm,height=5cm]{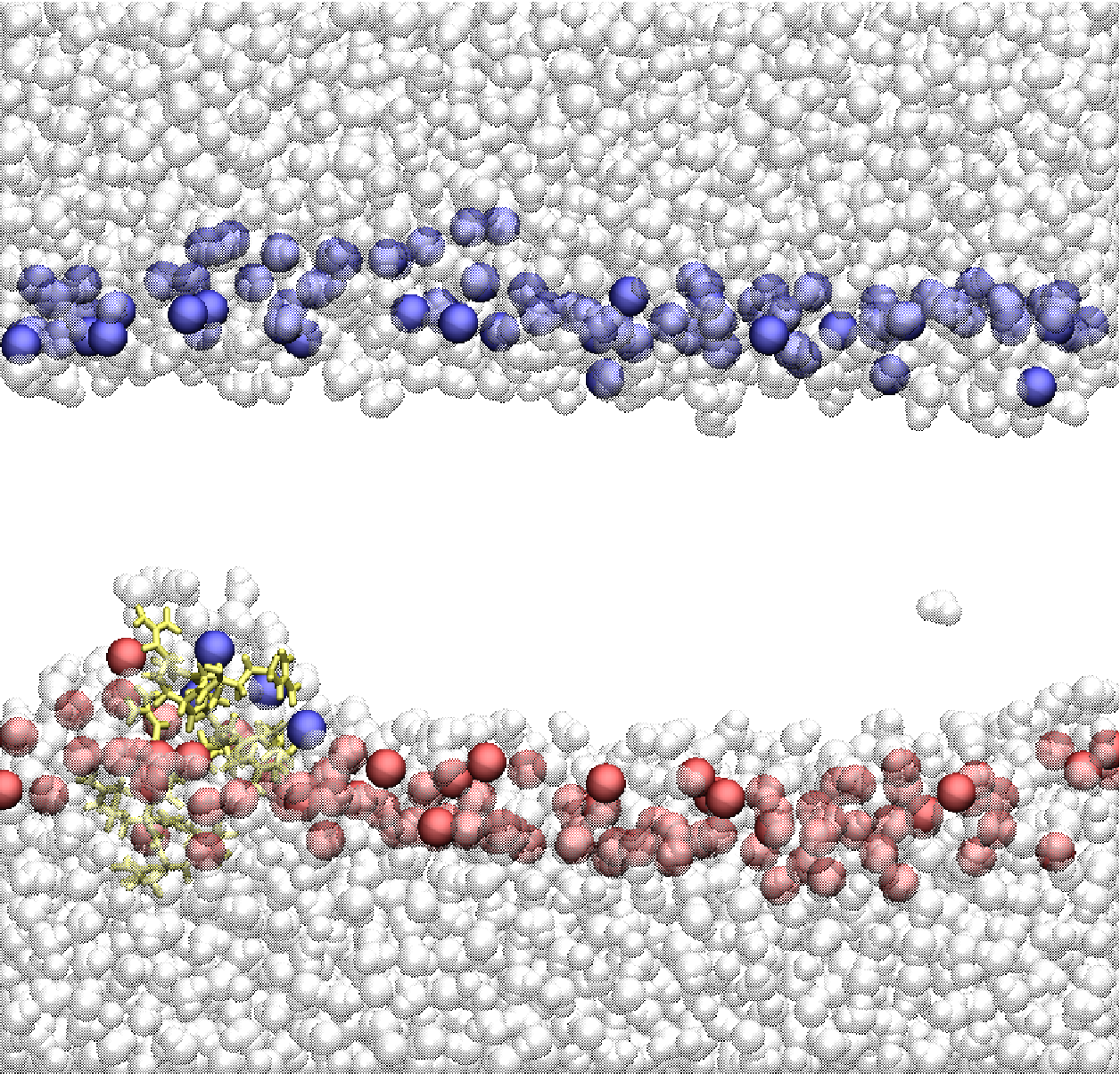}}\hfill
\subfloat[600 ns]{\includegraphics[width=5cm,height=5cm]{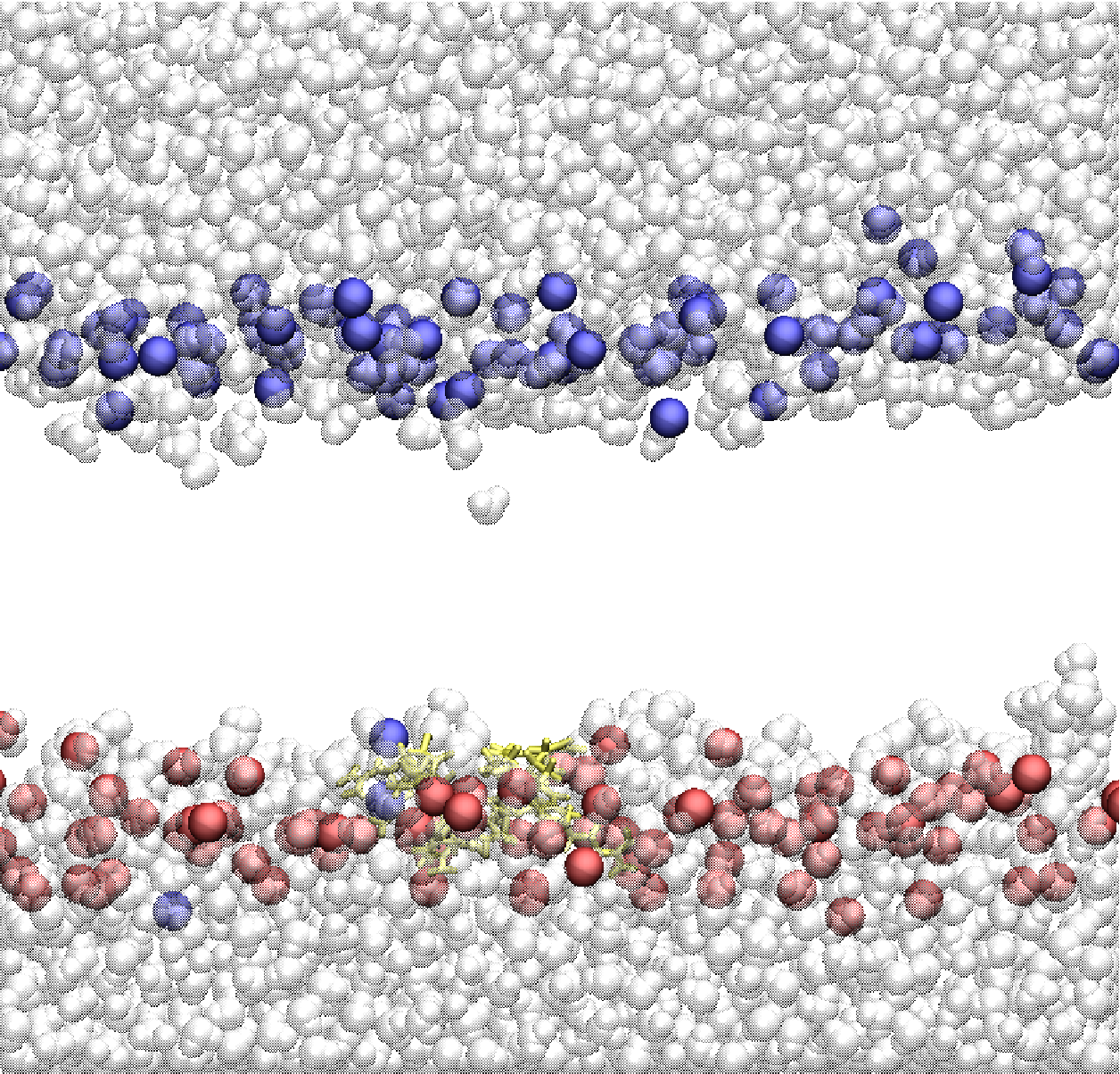}}\vspace{5mm}
\subfloat[800 ns]{\includegraphics[width=5cm,height=5cm]{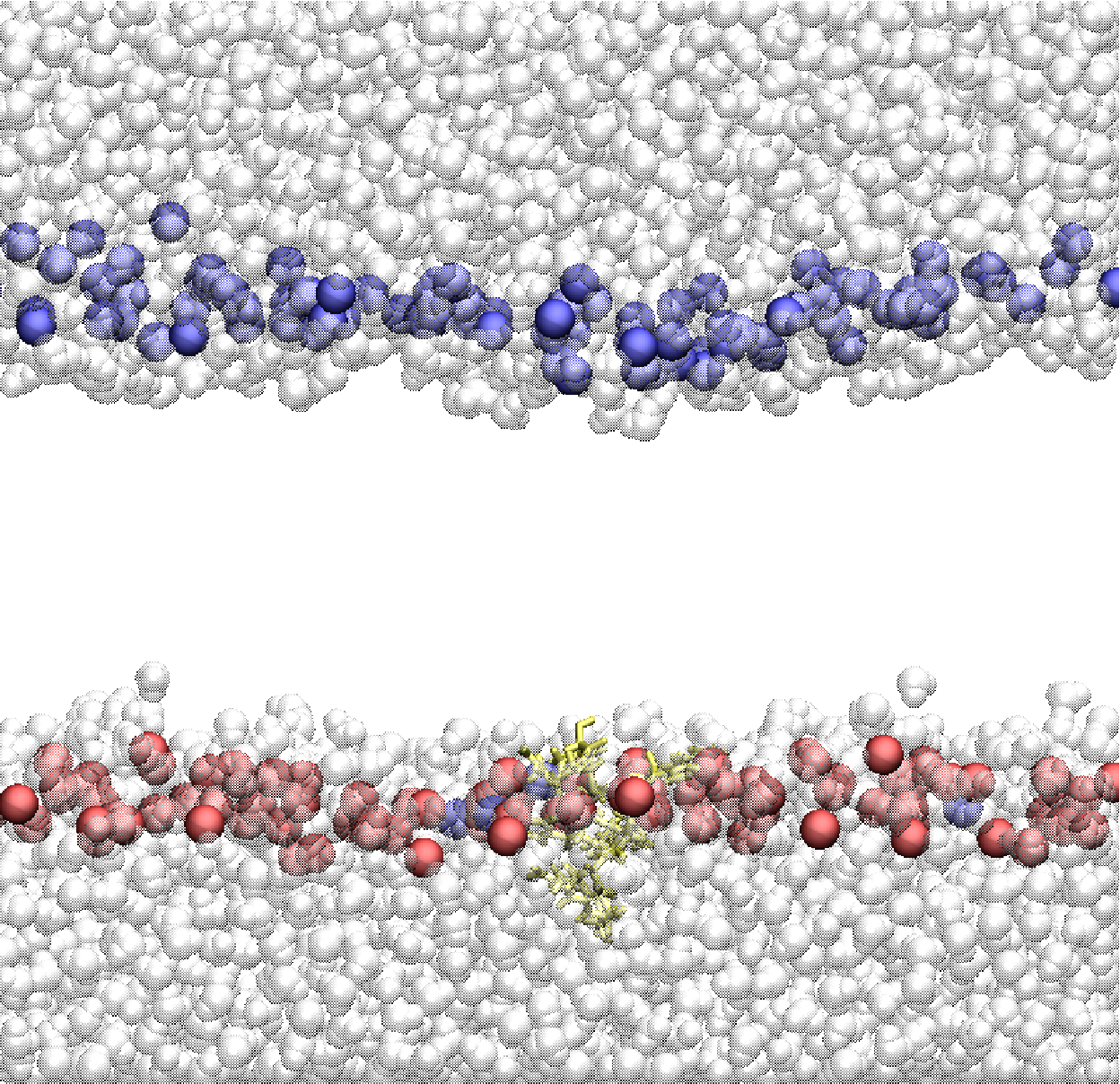}}\vspace{5mm}
\subfloat[1 $\mu$s]{\includegraphics[width=5cm,height=5cm]{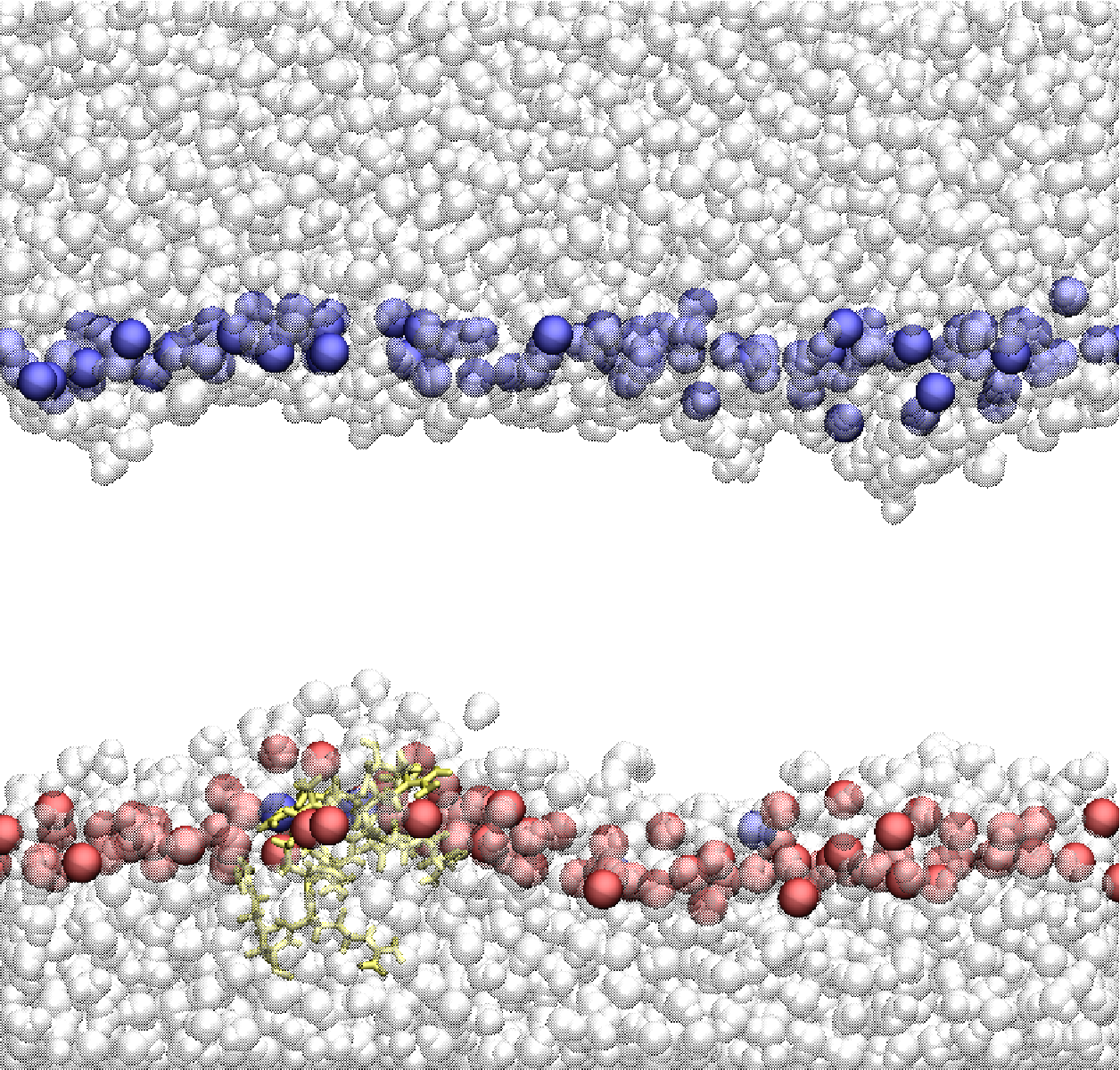}}

\caption{Snapshots at (a) 20 ns, (b) 40 ns, (c) 60 ns, (d) 600 ns, (e) 800 ns, and (f) 1 $\mu$s during the additional conventional MD simulation (yellow: Arg$_9$, white spheres: water, blue \& red : phosphorus atoms of the upper leaflet \& lower leaflet, respectively).}\label{figS4}
\end{figure}

\begin{figure}[ht!]
\centering
\includegraphics[width=10cm,height=10cm]{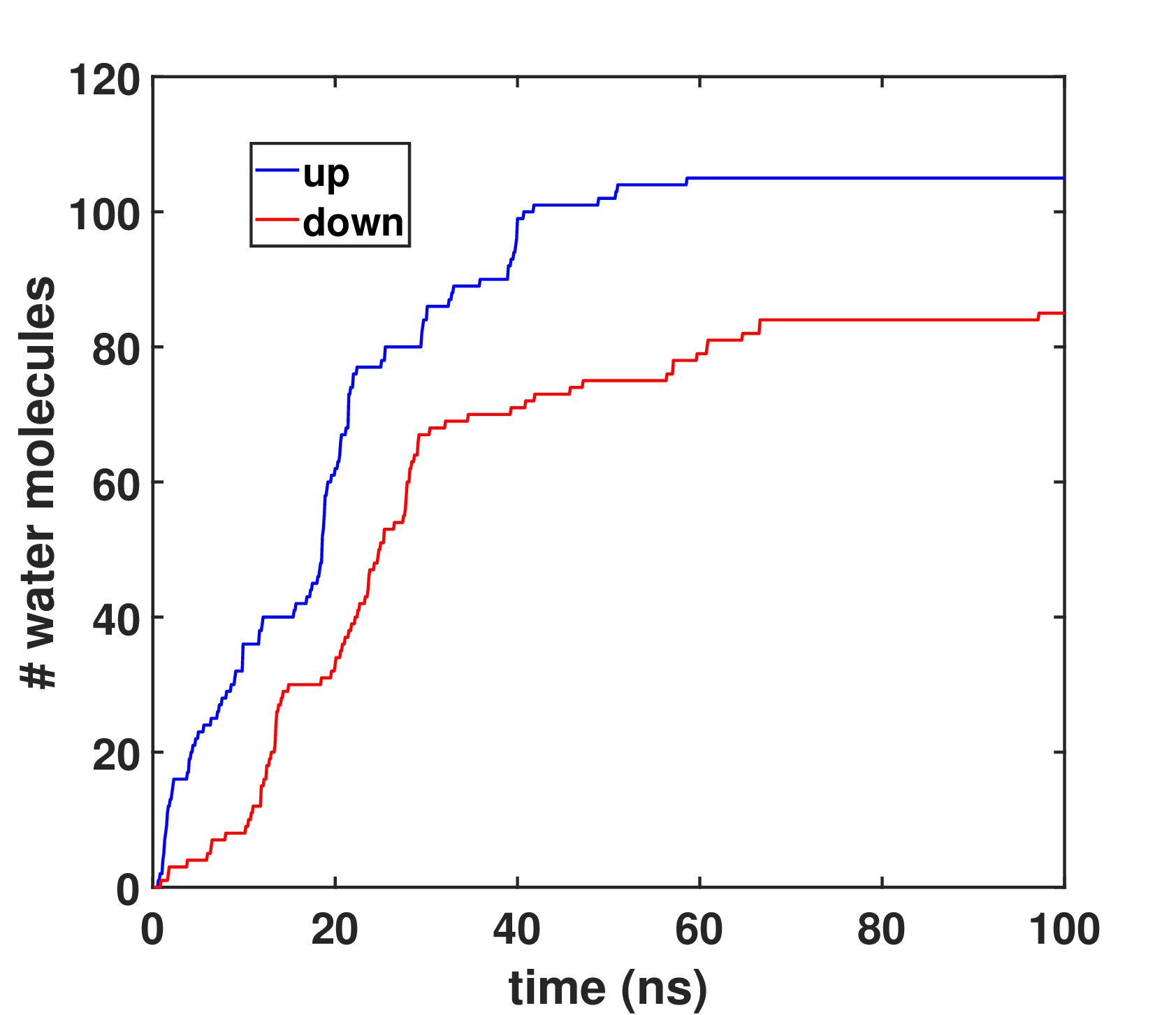}
\caption{The total accumulated number of water molecules translocated across the membrane (the moving-up(exiting) water \& the moving-down(entering) water) vs. the simulation time}\label{figS5}
\end{figure}

After finishing the WE simulations (from the WE1 to WE5), we ran an additional conventional MD simulation for 1 $\mu$s to see conformational changes of Arg$_9$s and the lipid molecules. The simulation method was similar to the previous ones\cite{Choe2020,Choe2021}. Fig. \ref{figS4} presents snapshots at a few different times during the simulation. During the simulation, Arg$_9$ moved back to the membrane's center with a few lipid molecules. The main text mentioned that six lipid molecules moved along with Arg$_9$ during the WE simulations. Four stayed in the lower leaflet until the end of the simulation, while two moved back to the upper leaflet when Arg$_9$ moved back to the center of the membrane. 
Since the simulation started, it took about 45 ns for two lipids to move back to the upper leaflet. The water flow and the movement of Arg$_9$ affect the phospholipid(PL) translocation. The water pore was closed after about 70 ns, as shown in Fig. \ref{figS5}. Fig. \ref{figS5} presents the total number of water molecules translocated across the membrane during the first 100 ns simulation. 
Before closing the water pore ($\sim$ 70 ns), the upward water flux was 1.5 (water molecules/ns), while the downward water flux was 1.2. The upward water flux is slightly higher than the downward water flux due to the upward movement of Arg$_9$.   

\bibliography{cpp}